\documentclass[prb,aps,twocolumn,amsmath,amssymb,floatfix,superscriptaddress,notitlepage]{revtex4-1}

\usepackage{lmodern}
\usepackage[T1]{fontenc}
\usepackage[english]{babel}
\usepackage{graphicx}
\usepackage[caption=false]{subfig}
\usepackage{amsmath}
\usepackage{mathrsfs}
\usepackage{amsfonts}
\usepackage{amsthm}
\usepackage{amssymb}
\usepackage{enumerate}
\usepackage{wrapfig}
\usepackage{mathtools}
\usepackage[colorlinks=true,linkcolor=blue]{hyperref}
\usepackage{xcolor}
\usepackage{physics}
\usepackage{listings}

\DeclareMathOperator{\sgn}{sgn}

\usepackage{stackengine}
\stackMath

\newcommand{\editR}[1]{\textcolor{black}{#1}}
\usepackage{comment}
\newcommand{\michel}[1]{{{\color{black}#1}}}
\newcommand{\new}[1]{{{\color{black}#1}}}

\begin{document}

\title{Thermodynamics of Non-Hermitian Josephson junctions with exceptional points}

\author{D. Michel Pino}
\affiliation{Instituto de Ciencia de Materiales de Madrid (ICMM), Consejo Superior de Investigaciones Científicas (CSIC), Sor Juana Inés de la Cruz 3, 28049 Madrid, Spain}
\author{Yigal Meir}
\affiliation{Department of Physics, Ben-Gurion University of the Negev, Beer-Sheva 84105, Israel}
\author{Ramón Aguado}
\email{ramon.aguado@csic.es}
\affiliation{Instituto de Ciencia de Materiales de Madrid (ICMM), Consejo Superior de Investigaciones Científicas (CSIC), Sor Juana Inés de la Cruz 3, 28049 Madrid, Spain}
\begin{abstract}
We present an analytical formulation of the thermodynamics, free energy and entropy, of any generic Bogoliubov de Genes model which develops exceptional point (EP) bifurcations in its complex spectrum when coupled to reservoirs. We apply our formalism to a non-Hermitian Josephson junction where, despite recent claims, the supercurrent does not exhibit any divergences at EPs. The entropy, on the contrary, shows a universal jump of $1/2\log 2$ which can be linked to the emergence of Majorana zero modes (MZMs) at EPs. Our method allows us to obtain precise analytical boundaries for the temperatures at which such Majorana entropy steps appear. We propose a generalized Maxwell relation linking supercurrents and entropy which could pave the way towards the direct experimental observation of such steps in e.g. quantum-dot-based minimal Kitaev chains.
\end{abstract}
\maketitle

\section{Introduction}

At weak coupling, an external environment only induces broadening and small shifts to the levels of a quantum system. In contrast, the strong coupling limit is highly nontrivial and gives rise to many interesting concepts in e.g. quantum dissipation \cite{Weiss}, quantum information science \cite{Harrington22} or quantum thermodynamics \cite{PhysRevLett.124.160601}, just to name a few. An interesting example is the emergence of spectral degeneracies in the complex spectrum (resulting from integrating out the environment), also known as exceptional point (EP) bifurcations, where eigenvalues and eigenvectors coalesce \cite{Berry2004}. During the last few years, a great deal of research is being developed in so-called non-Hermitian (NH) systems with EPs, in various contexts including open photonic systems \cite{Doppler2016}, Dirac \cite{Zhen2015}, Weyl \cite{Cerjan2019} and topological matter in general \cite{PhysRevLett.120.146402,PhysRevX.8.031079, PhysRevX.9.041015,RevModPhys.93.015005}.
\begin{figure}[ht]
\centering
    \includegraphics[width=\linewidth]{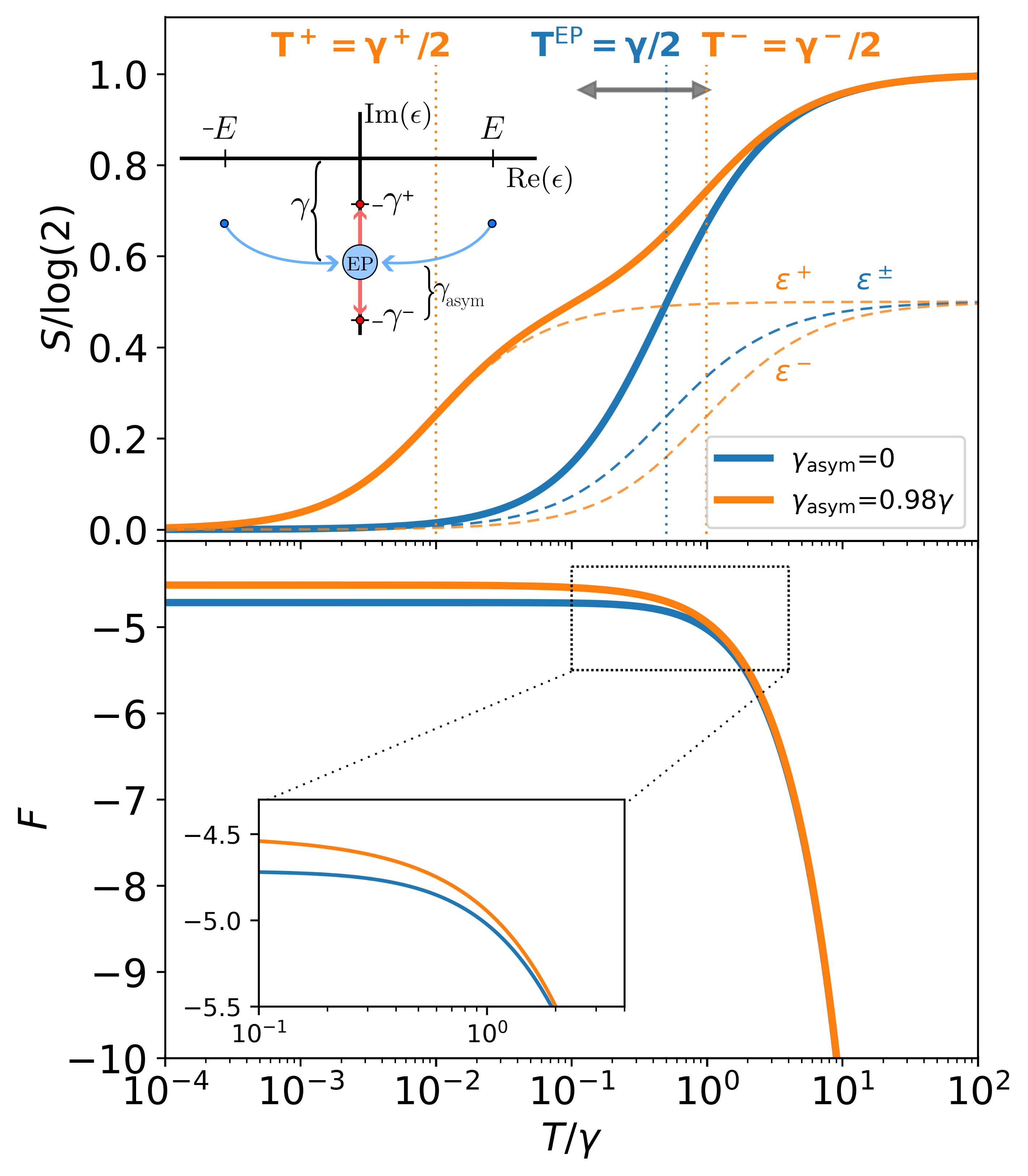}
    \caption{\textbf{Top (inset): Formation of an exceptional point}. The complex eigenvalues of a NH BdG Hamiltonian evolve as a function of some parameter until they coalesce at a so-called EP and then bifurcate into two purely imaginary eigenvalues with different decay rates to the reservoir $\gamma^\pm$, (quasi-bound MZMs). \textbf{Top (main): Entropy development after an exceptional point}. Before the EP, both eigenvalues have identical absolute values by particle-hole symmetry, $\epsilon^\pm=\pm E -i\gamma$, and thus both poles contribute to the entropy in \eqref{entropy} at the same temperature until the EP is reached ($T^{\mathrm{EP}}=|\epsilon^\pm|/2=\gamma/2$, blue curve). After the EP, $E=0$ and their imaginary parts are no longer identical, $\epsilon^\pm=-i\gamma^\pm$. Hence, their contributions to the entropy come at different temperatures ($T^{\pm}=\gamma^\pm/2$, orange curve), giving rise to a fractional plateau $S=\log(2)/2$ of width $\gamma_\mathrm{asym}=(\gamma^- - \gamma^+)/2$. \textbf{Bottom: Free energy} changes in the temperature dependence (by more than six decades) of $F$ are barely distinguishable before and after an EP, as opposed to the entropy above $S=-\partial F/\partial T$, which illustrates the subtlety of the calculations presented here. The inclusion of a finite cutoff leads to a convergent low-temperature asymptotic limit $F_{T\to 0}=\frac{1}{2}\sum_j\frac{\gamma_j}{\pi}(\log\frac{\gamma_j}{D}-1)$ see Eq. \eqref{limits} for $E_j=\mu=0$, which cures a divergence $F_{T\to 0}=\frac{1}{2}\sum_j\frac{\gamma_j}{\pi}(\log\frac{\gamma_j}{2\pi T}-1)$ (see Appendix \ref{sec:Appendix_B}).}
    \label{fig:EP-entropy}
\end{figure}
The role of NH physics and EP bifurcations
in systems with Bogoliubov-de Gennes (BdG) symmetry has hitherto remained unexplored, until recently. Specifically, 
there is an ongoing debate on how to correctly calculate the free energy in open BdG systems, a question relevant in e.g Josephson junctions coupled to external electron reservoirs. Depending on different approximations, such "NH junctions"  have been predicted to exhibit exotic effects including imaginary persistent currents \cite{PhysRevB.106.L121102,PhysRevB.103.035415} and supercurrents \cite{cayao2023nonhermitian,li2024anomalous} or various transport anomalies at EPs \cite{PhysRevLett.131.116001}. If one instead uses the biorthogonal basis associated with the NH problem \cite{shen2024nonhermitian}, or an extension of scattering theory to include external electron reservoirs  \cite{beenakker2024josephson}, the supercurrents are real and exhibit no
anomalies. At the heart of this debate is whether a straightforward use of the complex spectrum plugged into textbook definitions of thermodynamic functions leads to meaningful results or whether, on the contrary, NH physics needs to be treated with care when calculating the free energy. 

We here present a well-defined procedure, valid for arbitrary coupling and temperature, which allows us to calculate the free energy, Eq. \eqref{eq:free-energy}, without any divergences at EPs. Derivatives of this free energy, allow us to calculate
physical observables such as entropy \eqref{entropy} or supercurrents
 \eqref{supercurrent}. Interestingly, entropy changes of $\log2/2$ can be connected to emergent Majorana zero modes (MZMs) at EPs \cite{PhysRevB.87.235421,San-Jose2016,Avila_ComPhys2019}.
 While such fractional entropy steps were predicted before in seemingly different contexts \cite{Smirnov,SelaPRL:19,PhysRevLett.130.237002}, our analysis in terms of EPs allows us to obtain precise analytical boundaries for the temperatures at which they appear. We propose a novel Maxwell relation connecting supercurrents and entropy, Eq. \eqref{Maxwell}, which would allow the experimental detection of the effects predicted here.
\par

\section{Exceptional points in open BdG models}

The starting point of our analysis is the
description of an open quantum system in terms of a Green's function
\begin{equation}
\label{Geff}
G^\mathrm{eff}(\omega)=[\omega-H_\mathrm{eff}(\omega)]^{-1} \;,
\end{equation}
where $H_\mathrm{eff}(\omega)=H_Q+\Sigma^r(\omega)$ is an effective NH Hamiltonian which takes into account how the quantum system $H_Q$ is coupled to an external environment through the retarded self-energy $\Sigma^r(\omega)$. In what follows, we consider the case where an electron reservoir induces a tunneling rate \footnote{We consider the simplest case where this selfenergy originates from a tunneling coupling to an external electron reservoir in the so-called wideband approximation where both the tunneling amplitudes and the density of states in the reservoir are energy independent,  which allows to write  the tunneling couplings as constant \cite{PhysRevLett.68.2512}.}, such that the complex poles of Eq. \eqref{Geff} have a well-defined physical interpretation in terms of quasi-bound states.
If $H_\mathrm{eff}$ is a BdG Hamiltonian, electron-hole symmetry can be satisfied in two non-equivalent ways: (i) one can have pairs of poles with opposite real parts and with the same imaginary part $\epsilon^\pm=\pm E -i\gamma=-(\epsilon^{\mp})^*$; or, alternatively, (ii) two independent and purely imaginary poles $\epsilon^\pm=-i\gamma^\pm=-(\epsilon^{\pm})^*$. A bifurcation of the former, corresponding to standard finite-energy BdG modes with an equal decay to the reservoir $\gamma$, into the latter, two MZMs with different decay rates \cite{PhysRevB.87.235421,San-Jose2016,Avila_ComPhys2019}, defines an EP (Fig. \ref{fig:EP-entropy}a, inset).

\section{NH Free energy}

Calculating the free energy of an open quantum system is nontrivial since a direct substitution of a complex spectrum $\epsilon_j=E_j-i\gamma_j$ in the standard expression $F = -\frac{1}{\beta}\log Z=-\frac{1}{\beta}\sum_j \log \left( 1 + e^{-\beta(\epsilon_j-\mu)}\right)$, can lead to complex results and divergences after an EP 
\cite{cayao2023nonhermitian,li2024anomalous}. 
To avoid inconsistencies, one possibility is to use the occupation 
$\langle N\rangle =\int_{-\infty}^\infty d\omega G^< (\omega)=-\frac{\partial F}{\partial\mu}$, 
which is a well-defined quantity in open quantum systems, even in non-equilibrium situations where the so-called Keldysh lesser Green's function  $G^< (\omega)$ can be generalized beyond the fluctuation-dissipation theorem \cite{PhysRevLett.68.2512}.
In BdG language, $\langle N\rangle$ can be written as
\begin{equation}
    \langle N\rangle = \frac{1}{2}\int d\omega \, \Omega(D)\left[\rho^p(\omega) f(\omega-\mu) + \rho^h(\omega) f(-\omega-\mu)\right],
\label{occupation}
\end{equation}
where $f(\omega)$ is the Fermi-Dirac function and we have explicitly separated the total spectral function
\begin{equation}
\label{eq:rho}
    \rho(\omega) = -\frac{1}{2\pi}\Im\Tr G^\mathrm{eff}(\omega) = -\frac{1}{2\pi}\Im\sum_j\frac{1}{\omega-\epsilon_j} \;,
\end{equation}
in its particle ($\Re(\epsilon_j)>0$) and hole ($\Re(\epsilon_j)<0$) branches, $\rho^p(\omega)$ and $\rho^h(\omega)$, respectively \footnote{A global factor of $1/2$ arises from the duplicity of dimensions in BdG formalism.}. We have also added a Lorentzian cutoff $\Omega(D)=D^2/(D^2+\omega^2)$ to avoid divergences in the thermodynamic quantities as $T\to 0$ (see Appendix \ref{sec:Appendix_B}). The integral in Eq. \eqref{occupation} can be analytically solved by residues (Appendix \ref{sec:Appendix_B}) and then be used to obtain $F$ as\footnote{Note that while the above discussion is limited to non-interacting BdG Hamiltonians, the formalism can be generalized to an arbitrary interacting system. Assuming the number of particles is conserved, each eigenstate $\Psi_{N,j}$ has a specific particle number $N$ and energy $E_{N,j}$. The single-particle Green's function in Lehmann representation will then have a set of simple poles at $\pm(E_{N,j}-E_{(N-1),i})+i\eta$, that is, the difference between the energies of the states $\Psi_{N,j}$ and $\Psi_{(N+1),i}$. Adding to these poles a finite imaginary part coming from the coupling to a reservoir, as long as it only depends on the state, but not on the temperature $T$ or the chemical potential $\mu$, all the dependences with $T$ and $\mu$ will come from the Fermi function, just as for the non-interacting system.}
\begin{equation} \label{eq:free-energy}
\begin{aligned}
    F & = -\int d\mu \, \langle N\rangle= \frac{1}{2}\sum_j \left[ 2T\Re\log\Gamma\left(\frac{1}{2}+\frac{i\gamma_j-E_j+\mu}{2i\pi T}\right) \right.
    \\
    & \left. - \frac{2T\gamma_j}{D}\Re\log\Gamma\left(\frac{1}{2}+\frac{D+i\mu}{2\pi T}\right)  + h(T,E_j,\gamma_j) \right] \;,
\end{aligned}
\end{equation}
with $\log\Gamma(z)$ being the log-gamma function and $h(T,E_j,\gamma_j)$ a generic function coming from the integration. We now perform the limits $\gamma_j=0$ and $T\to 0$ of the previous expression,
\begin{equation}
\label{limits}
\begin{aligned}
    F_{\gamma=0} =& \frac{1}{2}\sum_j\left[ T\log(2\pi) - T\log\left(2\cosh\frac{E_j-\mu}{2T}\right) \right.
    \\
    & \left. - \frac{\mu}{2} + h(T,E_j,0) \right] \;,
    \\
    F_{T\to 0} =& \frac{1}{2}\sum_j \left[ \frac{\gamma_j}{\pi}\log\frac{\sqrt{\gamma_j^2+(E_j-\mu)^2}}{D} - \frac{\mu}{2} \right.
    \\
    & \left. - \frac{E_j-\mu}{\pi}\arctan\frac{E_j-\mu}{\gamma_j} + h(0,E_j,\gamma_j) \right] \;,
\end{aligned}
\end{equation}
which, by comparison with well-known limits \cite{zagoskin,Hewson}, give $h(T,E_j,\gamma_j)= - \gamma_j/\pi - T\log(2\pi)$
\footnote{Interestingly, this term is relevant for the entropy since it adds a physical contribution $\log(2\pi)$. In contrast, the term $-\gamma_j/\pi$ cancels out in the supercurrent (see Appendix \ref{sec:Appendix_B}).}. From now on, we fix $\mu=0$ but the complete derivation with full expressions, including $\mu\neq 0$, can be found in Appendix \ref{sec:Appendix_B}.
Derivatives of Eq. \eqref{eq:free-energy} allow us to obtain relevant thermodynamic quantities, as we discuss now.

\section{Entropy steps from EPs}

Using the free energy in Eq. \eqref{eq:free-energy}, the entropy, defined as $S=-\partial F/\partial T$, reads:
\begin{equation} \label{entropy}
\begin{aligned}
   S & = \frac{1}{2}\sum_j \left[ \log(2\pi) - 2\Re\log\Gamma\left(\frac{1}{2}+\frac{i\gamma_j-E_j}{2i\pi T}\right) \right.
   \\
   &  + \frac{2\gamma_j}{D}\log\Gamma\left(\frac{1}{2}+\frac{D}{2\pi T}\right) + \frac{\gamma_j}{\pi T}\Re\psi\left(\frac{1}{2}+\frac{i\gamma_j-E_j}{2i\pi T}\right)
   \\
   & \left. - \frac{E_j}{\pi T}\Im\psi\left(\frac{1}{2}+\frac{i\gamma_j-E_j}{2i\pi T}\right) - \frac{\gamma_j}{\pi T}\psi\left(\frac{1}{2}+\frac{D}{2\pi T}\right) \right] \;,
\end{aligned}
\end{equation}
where $\psi(z)$ is the digamma function. From Eq. \eqref{entropy}, we can define a crossover temperature $T_j=|\epsilon_j|/2$ as the inflection point when the eigenvalue $\epsilon_j$ begins to have a non-zero contribution ($S_j=\log(2)/4$) to the total entropy, Fig. \ref{fig:EP-entropy}a. Hence, two standard BdG poles $\epsilon^\pm = \pm E -i\gamma$ will contribute at the same temperature to the entropy of the system since their absolute values are equal, $|\epsilon^\pm|=\sqrt{E^2+\gamma^2}$, and thus a single plateau of $S=\log(2)$ can be measured. On the contrary, after an EP at $\epsilon^+=\epsilon^-=-i\gamma$, the poles bifurcate taking zero real parts and different imaginary parts, $\epsilon^+=-i\gamma^+$ and $\epsilon^-=-i\gamma^-$, and separating from each other a distance $\gamma^--\gamma^+$. Then, each pole will contribute to the entropy at a different temperature $T^\pm=|\epsilon^\pm|/2=\gamma^\pm/2$, giving rise to a fractional plateau $S=\log(2)/2$ of width $T^--T^+=(\gamma^--\gamma^+)/2$. Here it is very important to point out that this nontrivial behavior of the entropy seems absent in the free energy that we used for the calculation of  $S=-\partial F/\partial T$. Indeed,  $F$ is seemingly insensitive to any EP bifurcation even when varying the temperature over six decades, Fig. \ref{fig:EP-entropy}b. 

\section{Non-Hermitian Josephson junction with EPs}

Our method also allows to calculate the supercurrent of any generic NH Josephson junction (or similarly the persistent current through a normal ring \cite{shen2024nonhermitian}) by just considering a phase-dependent spectrum  $\epsilon_j(\phi)$  and taking phase derivatives of Eq. \eqref{eq:free-energy} as $I = \frac{\partial F}{\partial \Phi}=\frac{2e}{\hbar}\frac{\partial F}{\partial \phi}$, which gives
\begin{equation}
\label{supercurrent}
\begin{aligned}
    I(\phi) &= \frac{e}{\hbar}\sum_j\left[ -\frac{1}{\pi}\Im\psi\left(\frac{1}{2}+\frac{i\gamma_j-E_j}{2i\pi T}\right)\frac{\partial E_j}{\partial\phi} \right.
    \\
    & + \frac{1}{\pi}\Re\psi\left(\frac{1}{2}+\frac{i\gamma_j-E_j}{2i\pi T}\right)\frac{\partial \gamma_j}{\partial\phi} - \frac{1}{\pi}\frac{\partial \gamma_j}{\partial\phi}
    \\
    & \left. - \frac{2T}{D}\log\Gamma\left(\frac{1}{2} + \frac{D}{2\pi T}\right)\frac{\partial\gamma_j}{\partial\phi} \right] \;.
\end{aligned}
\end{equation}

\begin{figure}[h!]
    \centering
    \includegraphics[width=0.9\linewidth]{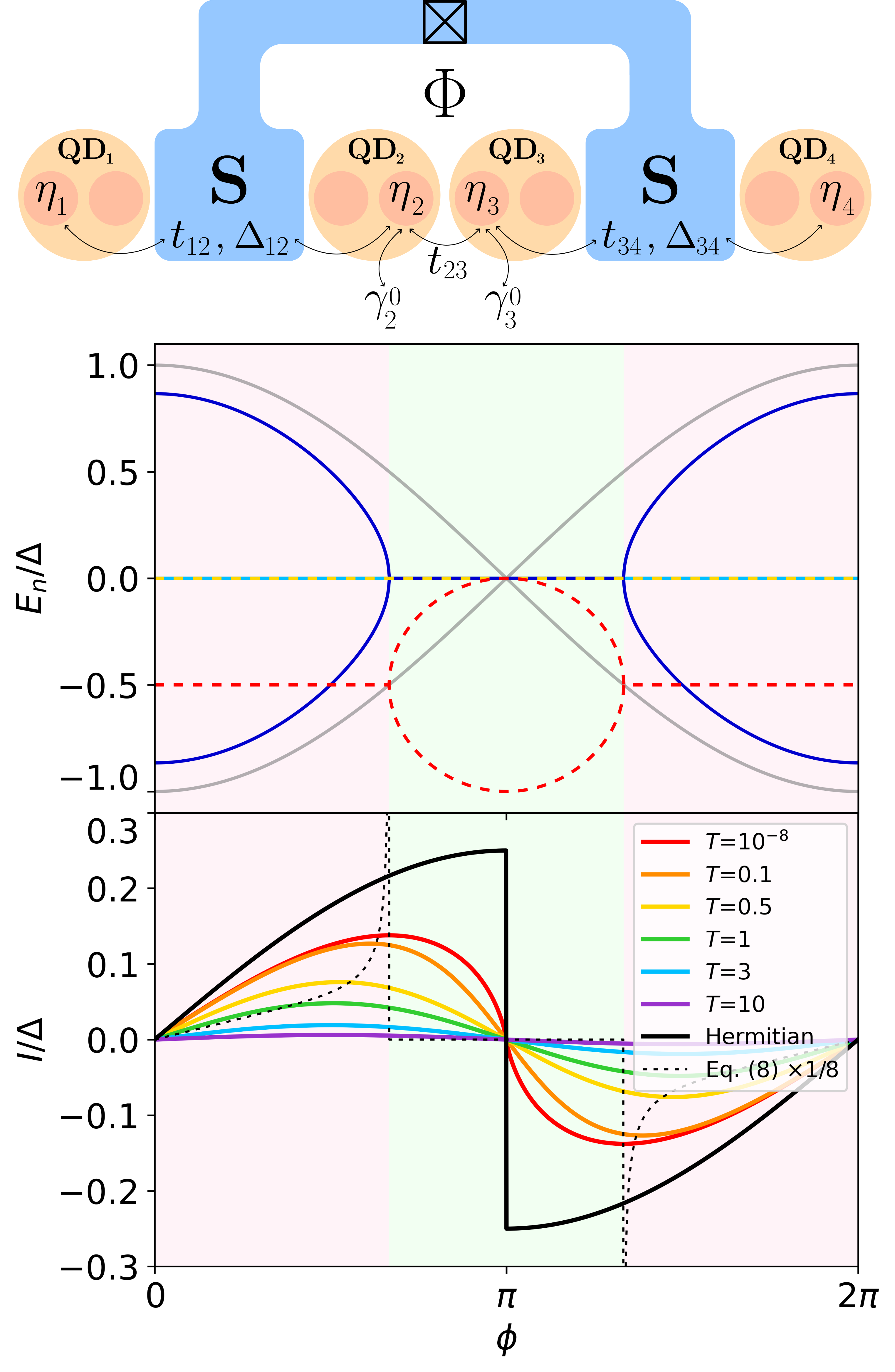}
    \caption{\textbf{Top: Schematic illustration of the four-Majorana Josephson junction device.} Each segment comprises two quantum dots connected via a middle superconductor in a so-called minimal Kitaev geometry. In the low-energy regime, only the Majorana modes located at the edges are considered in the model. The two inner modes are connected through a weak link, coupling $t_{23}$, which defines a Josephson junction. Its superconducting phase difference $\phi$ can be controlled by the magnetic flux $\Phi=\frac{\hbar}{2e}\phi$ through a SQUID loop connecting the superconductors, with the cross denoting an ancillary junction, see e.g. \cite{PhysRevLett.131.097001,Pita-Vidal2023}. The internal QDS are coupled to reservoirs with rates $\gamma^0_2\neq\gamma^0_3$. \textbf{Center: Energy spectrum of the junction showing an EP near phase $\phi=\pi$.} Lightblue/yellow (darkblue/red) lines correspond to real/imaginary parts of $\epsilon_\mathrm{outer}^\pm$ ($\epsilon_\mathrm{inner}^\pm$). Innermost states bifurcate around $\phi=\pi$, presenting two different topological phases (pink/green) divided by a pair of EPs. Gray lines correspond to the closed system ($\gamma_i^0=0$). \textbf{Bottom: Phase dependence of supercurrent.} Colored curves correspond to different temperatures (see legend). The black curve is associated with the Hermitian (closed) analog at $T=10^{-8}$, assuming equilibrium occupations (hence no $4\pi$ Josephson effect) which shows the typical sawtooth-like profile for a perfectly transparent Andreev level (see Appendix \ref{sec:Appendix_C}). The dashed line shows the calculation using the real part of Eq. \eqref{supercomplex}. System parameters are fixed as $t_{12}=\Delta_{12}$, $t_{34}=\Delta_{34}$, $t_{23}=\gamma^0_2=\Delta$ and $\gamma^0_3=0$.}
    \label{fig:panel2}
\end{figure}

As $T\to 0$, the supercurrent carried by a pair of BdG poles simply becomes \footnote{For a single pair of BdG poles, $\gamma$ is independent/dependent of the phase before/after the EP and the other way around for $E$ (see Appendix \ref{sec:Appendix_C}).}
\begin{equation} \label{eq:T-limit_supercurrent_ABS}
\begin{aligned}
    \text{before EP:} \quad I_{T\to 0} & = -\frac{e}{\hbar}\frac{2}{\pi}\arctan\left(\frac{E}{\gamma}\right) \frac{\partial E}{\partial\phi}\;,
    \\
    \text{after EP:} \quad I_{T\to 0} & = \frac{e}{2\hbar}\frac{2}{\pi}\log\left(\frac{\gamma^+}{\gamma^-}\right)\frac{\partial\gamma^+}{\partial\phi} \;,
\end{aligned}
\end{equation}
which, for example, allows us to calculate the supercurrent carried by Andreev levels in a short junction coupled to an electron reservoir almost straightforwardly (see Appendix \ref{sec:Appendix_C}). 
Note that, although $I(\phi)$ has a cutoff-dependent term, it cancels by the particle-hole symmetry of the problem (Appendix \ref{sec:Appendix_C}). Eqs. \eqref{eq:T-limit_supercurrent_ABS} \emph{strongly differ} from a calculation using directly the complex spectrum \cite{cayao2023nonhermitian,li2024anomalous}
\begin{equation}
\label{supercomplex}
    I^\mathrm{alt}(\phi)=\frac{e}{\hbar}\left(\frac{\partial E}{\partial\phi} - i\frac{\partial \gamma}{\partial\phi}\right) \;.
\end{equation}

Eqs. \eqref{eq:free-energy}, \eqref{entropy} and \eqref{supercurrent} are the main results of this paper and allow to calculate thermodynamics from generic open BdG models (arbitrary coupling and temperature) that can be written in terms of complex poles (Eq. \eqref{Geff}).

\section{Non-Hermitian minimal Kitaev Josephson junction}

As an application we now consider a quantum dot (QD) array in a so-called minimal Kitaev model
\begin{equation} \label{eq:QD_Hamiltonian}
H_{\mathrm{DQD}}  = -\sum_i\mu_{i} c_{i}^\dagger c_{i} 
 - t_{i,i+1}c_{i}^\dagger c_{i+1} + \Delta_{i,i+1} c_{i} c_{i+1}+\mbox{H.c.}\;,
\end{equation}
where $c_i^\dagger$ ($c_i$) denote creation (annihilation) operators on each QD with a chemical potential $\mu_{i}$. The QDs couple via a common superconductor that allows for crossed Andreev reflection and single-electron elastic co--tunneling, with coupling strengths $\Delta_{i,i+1}$ and $t_{i,i+1}$, respectively. 
Remarkably, only two QDs are enough to host two localized MZMs \footnote{
Here, it is important to clarify that MZMs in these minimal chains only appear in fine-tuned “sweet spots” in parameter space and
without topological protection, which is why they are often called
poor man’s Majorana modes. However, their non-local nature remains, where one MZM localizes in
each QD of the minimal chain. Interestingly, longer
chains, already at the three QD level, start to show a
bulk gap and some degree of protection \cite{bordin2024}} $\eta_1$ and $\eta_2$ when a so-called sweet spot is reached with $\Delta_{1,2}=t_{1,2}$. This theoretical prediction \cite{Leijnse} has recently been experimentally implemented~\cite{Dvir-Nature2023,haaf2024engineering}. Let consider now a second double QD array (Majoranas $\eta_3$ and $\eta_4$) that forms a Josephson junction with the former array with a coupling $H_{\mathrm{JJ}}=-t_{2,3}e^{i\frac{\phi}{2}}c_{2}^\dagger c_3+\mbox{H.c.}$, with $\phi$ being the superconducting phase difference between both arrays and 
$t_{2,3}$ the tunneling coupling between inner QDs. If, additionally, the two inner QDs are coupled to normal reservoirs with rates $\gamma_2^0$ and $\gamma_3^0$ this system is a realization of a Non-Hermitian Josephson junction containing Majorana modes (see the sketch in Fig. \ref{fig:panel2}a). 
In the low-energy regime, this model can be described in terms of four Majorana modes \cite{Pino}.
Assuming $\mu_i=0$ $\forall i$ and $\Delta_{12}=t_{12}$ and $\Delta_{34}=t_{34}$, only the inner Majoranas $\eta_2$ and $\eta_3$ are coupled and lead to BdG fermionic modes of energy $\epsilon_\mathrm{inner}^\pm = -\frac{i}{2}\gamma_0 \pm \frac{1}{2}\Lambda(\phi)$, where $\gamma_0=\gamma_2^0+\gamma_3^0$, $\delta_0=\gamma_2^0-\gamma_3^0$ and $\Lambda(\phi)=\sqrt{2t_{23}^2(1+\cos\phi)-\delta_0^2}$, while the outer modes remain completely decoupled and $\epsilon_\mathrm{outer}^\pm=0$. For $\delta_0= 0$, one recovers the standard Majorana Josephson term: $\epsilon_\mathrm{inner}^\pm = -\frac{i}{2}\gamma_0 \pm t_{23}\cos(\phi/2)$. 
When $\delta_0\neq 0$, the spectrum develops EPs at phases $\phi_\mathrm{EP} =\arccos{[\frac{\delta_0^2}{2t_{23}^2}-1]}$. For an example of the resulting supercurrents and a comparison against Eq. \eqref{supercomplex}, see Fig. \ref{fig:panel2}.
Similarly, when $\Delta_{12}=\Delta_{34}=\Delta$, $t_{12}=t_{34}=t$ but $t\neq\Delta$, an additional pair of EPs appears at \footnote{Analytical expressions of eigenvalues, a similar calculation for a junction containing Andreev levels, as well as detailed discussions can be found in the Appendix \ref{sec:Appendix_C}.} \begin{equation}
\phi_\mathrm{EP} = \arccos{[\frac{\delta_0^2 - [4(\Delta-t) - \gamma_0]^2}{2t_{23}^2} - 1]}.
\label{eq:EP_4M-decoupled}
\end{equation}

\begin{figure}[ht]
    \centering
    \includegraphics[width=0.9\columnwidth]{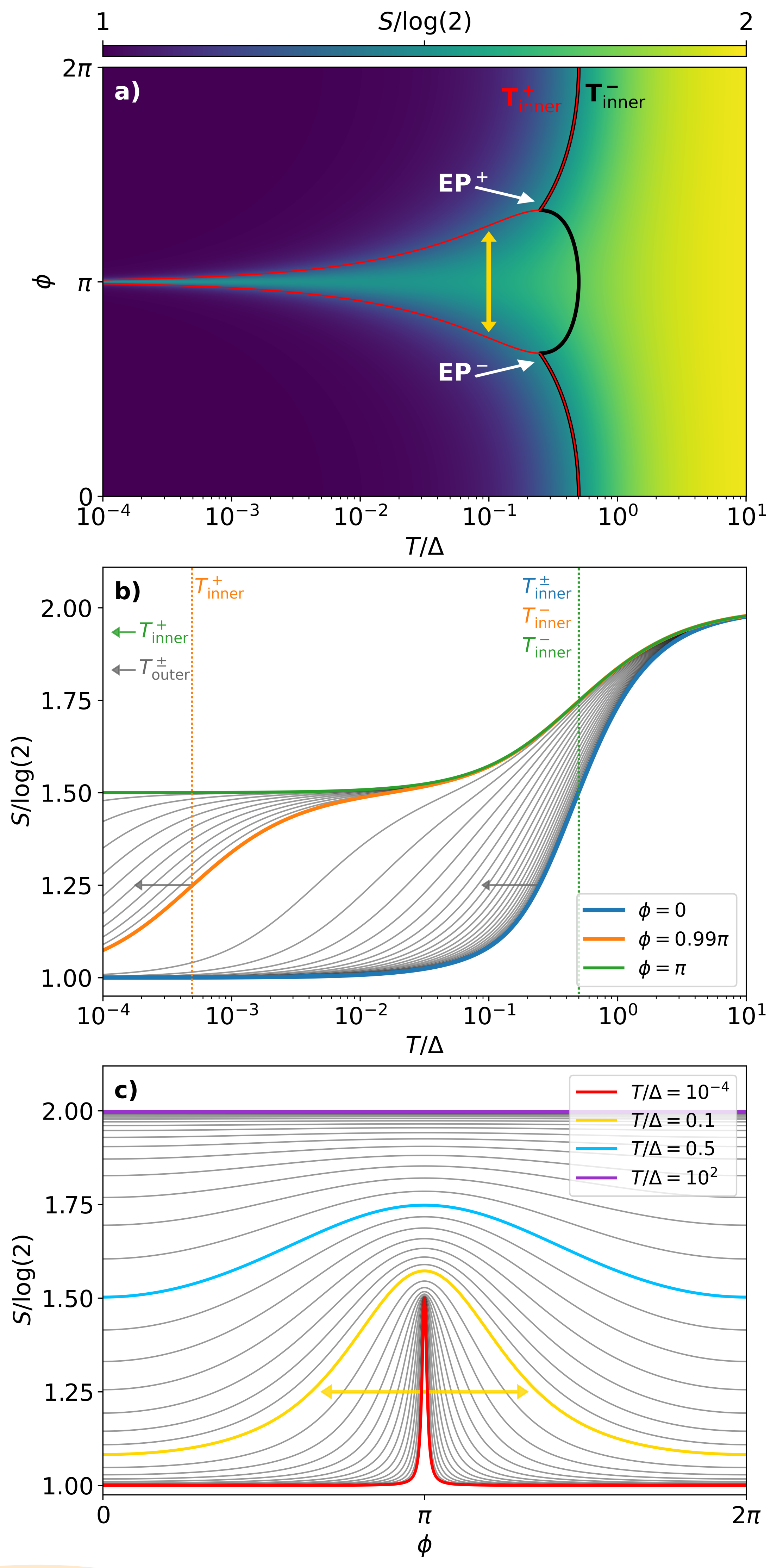}
    \caption{\textbf{Entropy of a Majorana Josephson junction with EPs.} \textbf{(a)} Entropy as a function of $\phi$ and $T$. Two EPs$^\pm$ are marked in white at $\phi_\mathrm{EP}$ and $2\pi-\phi_\mathrm{EP}$ (\ref{eq:EP_4M-decoupled}). Black/red curves correspond to $T_\mathrm{inner}^\pm$, agreeing with the jumps in entropy. \textbf{(b)} Entropy as a function of $T$ for different phases. We have also marked the position of $T_\mathrm{inner}^\pm$ for each curve. \textbf{(c)} Entropy as a function of $\phi$ for different temperatures. Gray intermediate steps follow the trajectory in between such colored curves. System parameters are fixed as $t_{12}=\Delta_{12}$, $t_{34}=\Delta_{34}$, $t_{23}=\gamma^0_2=\Delta$ and $\gamma^0_3=0$.}
    \label{fig:panel3}
\end{figure}

Using these analytics, the crossover temperatures  $T_j=|\epsilon_j|/2$ read
\new{
\begin{equation}
\label{TcriticialKitaev}
\begin{aligned}
    &\text{before EP: }\; T^\pm_\mathrm{inner} = \frac{1}{4}\sqrt{\gamma_0^2 + \Lambda^2(\phi)} \;,
    \\
    &\text{after EP: } \left\{ \begin{aligned}
        & T^+_\mathrm{inner} = \frac{\gamma_0-\sqrt{\delta_0^2-2t_{23}^2(1+\cos\phi)}}{4}
        \\
        & T^-_\mathrm{inner} = \frac{\gamma_0+\sqrt{\delta_0^2-2t_{23}^2(1+\cos\phi)}}{4}
    \end{aligned} \right.
\end{aligned}
\end{equation}
}
To illustrate their physical meaning, we plot
a full calculation of the entropy $S(T,\phi)$ using Eq. \eqref{entropy}, Fig. \ref{fig:panel3}a, together with the analytical expressions in Eq. \eqref{TcriticialKitaev} (solid lines). This plot demonstrates that changes in entropy can be understood from EPs, a claim that is even clearer by analyzing 
cuts at fixed phase (Fig. \ref{fig:panel3}b). Interestingly, a universal entropy change of $1/2\log 2$ can be linked to the emergence of MZMs at phase $\phi=\pi$ as $T\to 0$.
Alternatively, the entropy loss due to Majoranas can be seen in phase-dependent cuts taken at different temperatures, Fig. \ref{fig:panel3}c, which show as an interesting behavior where a $S=2\log 2$ plateau at large temperatures becomes an emergent narrow resonance, centered at $\phi=\pi$ and of height $S=1.5\log 2$, as $T$ is lowered.

\section{Temperature effects}

While in the minimal Kitaev chain model in Eq. \eqref{eq:QD_Hamiltonian} the pairing potential $\Delta$ is assumed to be a temperature-independent parameter, in a real experiment it comes from crossed Andreev reflections mediated by a middle segment separating both quantum dots. Such segment is typically a semiconducting region proximitized by a superconductor such that the parent superconducting gap  $\Delta_0\gg\Delta$  (in the experiments of Ref. ~\cite{Dvir-Nature2023,haaf2024engineering}, the gap of the parent aluminum superconductor is $\Delta_0\approx 270\mu$eV $\gg\Delta\approx 10\mu$eV). Thus, a regime where $T\gg\Delta$ without destroying superconductivity is possible. 
In this section, we include the BCS-like temperature dependence of $\Delta_0$ in a microscopic model of a minimal Kitaev chain to fully support this argument.

The DQD model of Eq. \eqref{eq:QD_Hamiltonian} in the main text can be realized microscopically by means of a common superconductor-semiconductor hybrid region which couples both QDss. Specifically, they couple via a subgap Andreev bound state (ABS) living in the middle region. Such ABS coupling allows for crossed Andreev reflection (CAR) and single-electron elastic co-tunneling (ECT), with coupling strengths $\Delta$ and $t$, respectively. Physically, the spin-orbit coupling in semiconductor-superconductor provides a spin-mixing term for tunneling electrons. The spin-mixing term can lead to finite ECT and CAR amplitudes for spin-polarized QDs when the spin-orbit and magnetic fields are non-colinear.  This ABS provides a low-excitation energy for ECT and CAR, which, starting from a microscopic model of two QDs connected by a semiconductor-superconductor
middle region, can be obtained from a Schrieffer-Wolff transformation to obtain effective cotunneling-like terms \cite{PhysRevLett.129.267701}
\begin{equation}
\label{CAR-ECT}
\begin{aligned}
    t & \sim \frac{t_L t_R}{\Delta_0} \left(\frac{2uv}{E_\mathrm{ABS}/\Delta_0}\right)^2
    \\
    \Delta & \sim \frac{t_L t_R}{\Delta_0} \left(\frac{u^2-v^2}{E_\mathrm{ABS}/\Delta_0}\right)^2
\end{aligned}
\end{equation}
where $t_{L/R}$ are the local tunneling strengths between each QD and the central region and
\begin{equation}
    E_\mathrm{ABS}=\Delta_0\sqrt{z^2 + 1}
\end{equation}
is the energy of the Andreev state, with $z\equiv\mu_\mathrm{ABS}/\Delta_0$ being the chemical potential of the middle region, and BdG coefficients given by $u^2=1-v^2=1/2 + \mu_\mathrm{ABS}/(2E_\mathrm{ABS})$. As discussed in Ref.~\cite{PhysRevLett.129.267701}, this energy dependence of $\Delta$ and $t$ allows one to vary their relative amplitudes by changing the chemical potential $\mu_\mathrm{ABS}$ of the ABS and tune them to precise points, so-called "sweet spots", where $\Delta=t$, a precise tuning \emph{which has already been demonstrated in different experimental platforms and configurations} \cite{Dvir-Nature2023,haaf2024engineering,haaf2024edgebulkstatesthreesite}. 

In Eqs. (\ref{CAR-ECT}), the gap $\Delta_0$ of the parent superconductor, is assumed to be temperature-independent \cite{PhysRevLett.129.267701}. We now extend this microscopic theory and explicitly consider the BCS-like temperature dependence of the Aluminium parent gap $\Delta_0\rightarrow \Delta_T$ which makes the crossed Andreev reflection and elastic cotunneling terms in Eq. \eqref{CAR-ECT} temperature-dependent too.
Interestingly, their temperature dependence is highly nonmonotonic, with large regions where they remain nearly constant until they decrease near the Aluminium critical temperature $T_c\approx 141.651  \; \mu\mathrm{eV}$ where the parent gap closes (see Appendix \ref{sec:Appendix_F}). 

\begin{figure}[ht]
    \centering
    \includegraphics[width=\linewidth]{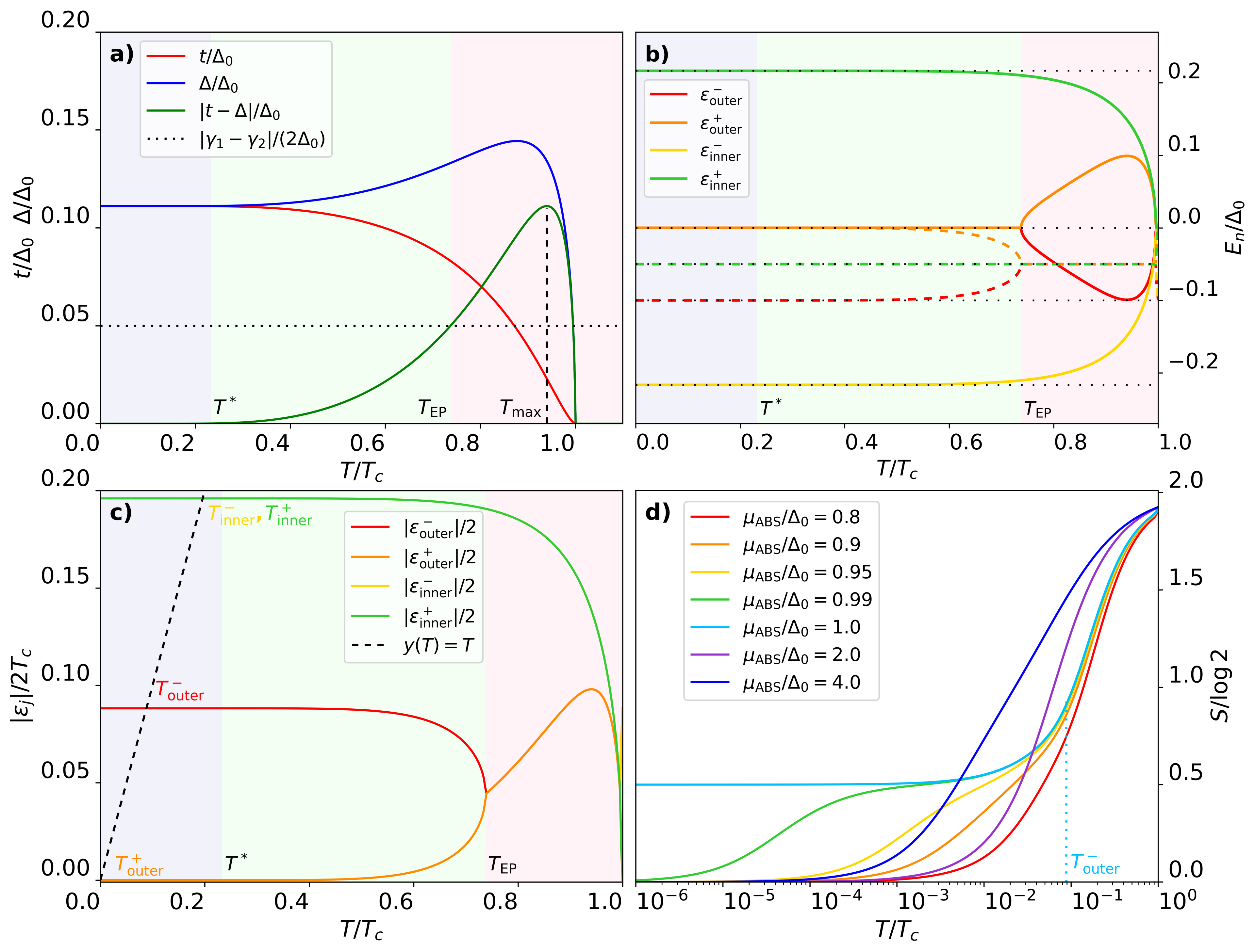}
    \caption{\textbf{(a) Temperature dependence of $t$ and $\Delta$}. Lilac/green regions border corresponds to $T^*$, such that $|t-\Delta|=10^{-4}\Delta_0$. Green/pink regions border corresponds to $T_\mathrm{EP}$, where appears the EP bifurcation between outer states ($|t-\Delta|=|\gamma_1^0-\gamma_2^0|/2$). The maximum value of $|t-\Delta|$ is marked with a vertical dashed line at $T_\mathrm{max}$. \textbf{(b) Complex energy spectrum of the system}. Solid/dashed lines correspond to the real/imaginary parts of $\epsilon_\mathrm{inner/outer}^\pm$.
    Black dotted lines correspond to the $T\to 0$ approximation of Eq. (\ref{eq:T-0_approx}), which, for this choice of parameters, agree with $T$-finite results until $T^*\approx 0.23T_c$ (lilac region in both figures defined as $|t-\Delta|\leq 10^{-4}\Delta_0$).
    \textbf{(c) Crossover temperatures}. Temperature dependence of the absolute value of the complex poles for $\mu_\mathrm{ABS}=\Delta_0$. Colored stars mark the points where $T=|\epsilon_j(T)|/2$, corresponding to the crossover temperatures of each pole. \textbf{(d) Entropy}. Temperature dependence of entropy for different values of $\mu_\mathrm{ABS}$. The crossover temperature $T_\mathrm{outer}^-$ is marked for the case $\mu_\mathrm{ABS}=\Delta_0$. Parameters used: $\gamma_1^0=0$ and $\gamma_2^0=0.1\Delta_0$; $\mu_\mathrm{ABS}/\Delta_0=1$ unless otherwise stated; $t_L=t_R=2/3\Delta_0$.}
    \label{fig:End-Matter}
\end{figure}
Specifically, we study the temperature dependence of the difference $|t-\Delta|$ (Fig. \ref{fig:End-Matter}a) for $\mu_\mathrm{ABS}=\Delta_0$, which corresponds to the sweet spot at zero temperature. If, additionally, each QD is coupled to normal reservoirs with rates $\gamma_1^0\neq\gamma_2^0$, this temperature evolution governs the emergence of exceptional point (EP) bifurcations, which are now temperature-dependent (Fig. \ref{fig:End-Matter}b). As the temperature increases in Fig. \ref{fig:End-Matter}a, both couplings remain equal $|t-\Delta|=0$ until reaching a temperature $T^*$ (lilac regions). Above $T^*$ both couplings start to differ $|t-\Delta|\neq 0$.
Nevertheless, as long as the condition $|t-\Delta|<|\gamma_1^0-\gamma_2^0|/2$ is fulfilled, the system still hosts a pair of MZMs (green region). This can be explicitly seen in Fig. \ref{fig:End-Matter}b where we plot the full temperature-dependent evolution of the complex spectrum. Specifically, the green region corresponds to a regime where two MZMs ($\Re(\epsilon_\mathrm{outer}^+)=\Re(\epsilon_\mathrm{outer}^-)=0$) are asymmetrically coupled to the reservoir ($\Im(\epsilon_\mathrm{outer}^+)\neq \Im(\epsilon_\mathrm{outer}^-)\neq 0$). This finite-temperature regime \emph{with MZMs} persists until $|t-\Delta|$ is large enough to close the bifurcation. Specifically, the exact value of the temperature at which the EP forms, $T_\mathrm{EP}$, is determined by the point where the condition $|t-\Delta|=|\gamma_1^0-\gamma_2^0|/2$ is fulfilled
\begin{equation}
    |t-\Delta| = \frac{t_Lt_R\Delta_T|\Delta_T^2-\mu_\mathrm{ABS}^2|}{(\Delta_T^2+\mu_\mathrm{ABS}^2)^2}= \frac{|\gamma_1^0-\gamma_2^0|}{2} \;.
\end{equation}
Above $T_\mathrm{EP}$, namely $|t-\Delta|>|\gamma_1^0-\gamma_2^0|/2$ the system no longer contains MZMs (pink region). The lilac region, on the other hand, should be understood as the temperature $T^*$ below which the $T\to 0$ limit is valid 
\begin{eqnarray}\label{eq:T-0_approx}
\begin{aligned}
   && \epsilon_\mathrm{outer}^\pm = -i\frac{\gamma_1^0+\gamma_2^0}{2} \pm i \frac{|\gamma_1^0-\gamma_2^0|}{2} = -i\gamma_{1/2}^0
   \quad,\quad \\
  && \epsilon_\mathrm{inner}^\pm = -i\frac{\gamma_1^0+\gamma_2^0}{2} \pm \sqrt{\frac{t_L^2t_R^2}{4\Delta_0^2} - \frac{(\gamma_1^0-\gamma_2^0)^2}{4}} \;.
\end{aligned}
\end{eqnarray}

Even though the calculations in Fig. \ref{fig:End-Matter} have been performed for the particular (but reasonable) values of $t_{L/R}=2/3\Delta_0$ and $|\gamma_1^0-\gamma_2^0|=0.1\Delta_0$, we can extract some general conclusions that enable the possibility of an experimental demonstration of our claim: although the magnitude of $|t-\Delta|$ depends on the local tunneling strengths $t_{L/R}$, their functional form against the temperature remains the same, reaching a maximum value 
\begin{equation}
    |t-\Delta|_\mathrm{max} = \frac{t_Lt_R}{4\mu_\mathrm{ABS}}
\end{equation}
at $\Delta_T=(\sqrt{2}-1)\mu_\mathrm{ABS}$, which is independent of $\gamma_{1/2}^0$ and $t_{L/R}$, and corresponds to a large value of $T_\mathrm{max}\approx 0.94T_c$ when $\mu_\mathrm{ABS}=\Delta_0$. Conversely, $T_\mathrm{EP}$ depends on the ratio between the reservoir coupling asymmetry $|\gamma_1^0-\gamma_2^0|$ and the tunneling strengths $t_{L/R}$, and it must be smaller than $T_\mathrm{max}$ such that the EP can develop. For the choice of parameters in Fig. \ref{fig:End-Matter}, $T_\mathrm{EP}\approx 0.74T_c<T_\mathrm{max}$. Particularly, for values of $|\gamma_1^0-\gamma_2^0|/2$ greater than $|t-\Delta|_\mathrm{max}=\frac{t_Lt_R}{4\Delta_0}$ ($\mu_\mathrm{ABS}=\Delta_0$), the system would never leave the non-trivial topological phase. By rewriting the couplings as $t_L=\eta_L\Delta_0$ and $t_R=\eta_R\Delta_0$, this condition reads $|\gamma_1^0-\gamma_2^0|>\eta_L\eta_R\Delta_0/2$. Since in realistic settings both $\eta_L,\eta_R\ll 1$, this implies that there is no need of a huge coupling asymmetry for having MZMs.

We now calculate the entropy of a minimal Kitaev chain (Fig. \ref{fig:End-Matter}d) including the temperature dependence of the effective parameters $\Delta$ and $t$, and hence of the eigenvalues that bifurcate. 
By tuning $\mu_\mathrm{ABS}=\Delta_0$, the system approaches the sweet-spot regime for temperatures below $T_\mathrm{EP}$, where the poles $\epsilon_\mathrm{outer}^\pm$ exhibit an EP (Fig. \ref{fig:End-Matter}b). 
On the other hand, as explained in the main text, the entropy jumps associated with these poles occur at the crossover temperatures $T_\mathrm{outer}^\pm = |\epsilon_\mathrm{outer}^\pm|/2$. If the magnitudes of the poles differ, $|\epsilon_\mathrm{outer}^-|\neq|\epsilon_\mathrm{outer}^+|$, a fractional plateau $S = \log(2)/2$ appears, and its width is given by $T_\mathrm{outer}^- - T_\mathrm{outer}^+$. Thus, this fractional plateau is observable for temperatures below $T_\mathrm{outer}^-$. Importantly, this temperature is not directly given by $T_\mathrm{EP}$: although the system hosts MZMs up to $T_\mathrm{EP}$, the fractional plateau only emerges when the mode most coupled to the reservoir has not yet contributed to the entropy, which happens at $T_\mathrm{outer}^-$. Here, it is important to emphasize that since $\epsilon_\mathrm{outer}^-(T)$ depends on the temperature (Fig. \ref{fig:End-Matter}b), the value of this crossover temperature becomes a self-consistent problem,
\begin{equation}
    T_\mathrm{outer}^- = \frac{|\epsilon_\mathrm{outer}^-(T_\mathrm{outer}^-)|}{2}.
\end{equation}
Thus, $T_\mathrm{outer}^-$ will be given by the crossing point between the curve $|\epsilon_\mathrm{outer}^-(T)|/2$ and the line $y(T)=T$ (and the same for the rest of poles), see Fig. \ref{fig:End-Matter}c. Importantly, in this figure all the crossover temperatures lie in the lilac region, where the limit $T\to 0$ is valid. 
Under this limit, we have in general $T_\mathrm{outer}^-=\max(\gamma_1^0,\gamma_2^0)/2$, being completely valid as long as $\gamma_1^0,\gamma_2^0<2T^*$.

Specifically, for the particular choice of parameters in Fig. \ref{fig:End-Matter} (cyan line of panel d for $\mu_\mathrm{ABS}=\Delta_0$), this temperature is one order of magnitude smaller than $T_c$ ($T_\mathrm{outer}^-=0.088T_c<T^*$), so that the $T\to 0$ approximation is valid for this and the rest of the poles, Fig. \ref{fig:End-Matter}c. Using the Aluminium $T_c$ this gives a temperature $T_\mathrm{outer}^-\approx 15\mu\mathrm{eV}$ (which in real temperature units corresponds to a temperature of $T_\mathrm{outer}^-\approx 170mK$ well within the experimental range of observability) where the fractional plateau (cyan line for $\mu_\mathrm{ABS}=\Delta_0$) develops. Importantly, in the experimental demonstrations of minimal Kitaev chains based on quantum dots ~\cite{Dvir-Nature2023,haaf2024engineering,haaf2024edgebulkstatesthreesite} $\Delta\approx 10-20\mu$eV, in the same energy range of our estimation of the temperature range below which the fractional plateau in the entropy emerges. This supports our claim that unreasonably low temperatures $T\ll \Delta$ are \emph{not needed} to observe our predicted fractional plateaus.
 
\section{Experimental detection of fractional entropy}

Recently it has been demonstrated that one can measure entropies of mesoscopic systems, either via Maxwell relations \cite{NatPhys.14.1083, PhysRevLett.129.227702} or via thermopower \cite{NatCommun.10.5801, nanolett.1c03591}. The Maxwell relation method relies on continuously changing a parameter $x$ (e.g. chemical potential or magnetic field) while measuring its conjugate variable $y$ (e.g. electron number or magnetization, respectively) such that $y=\partial F/\partial x$. Then, the Maxwell relation yields $\partial S/\partial x = -\partial y/\partial T$. Here we propose a novel application of this procedure, employing the Josephson current $I(\phi)$, which gives
\new{
\begin{equation}
\label{Maxwell}
    S(\phi_2) - S(\phi_1) = -\int_{\phi_1}^{\phi_2}\frac{\partial I(\phi)}{\partial T}\, d\phi \;.
\end{equation}
}
Since the phase difference on the Josephson junction can be controlled (Fig. \ref{fig:panel2}a) by e.g. embedding it in a SQUID loop \cite{PhysRevLett.131.097001,Pita-Vidal2023}, one can integrate $dI/dT$ between $\phi_1=0$ and $\phi_2=\pi$. From Fig.~\ref{fig:panel3}b we expect $\Delta S$ to change from zero at high-T to $\log2/2$ at low-T, an unequivocal signature of MZMs in the junction. 

\new{Our estimates using a microscopic model of a minimal Kitaev chain based on two QDs coupled through an Andreev bound state localized in a hybrid semiconductor-superconductor segment \cite{PhysRevLett.129.267701}, and including the temperature dependence of the effective parameters for cross Andreev reflection ($\Delta$) and single-electron elastic cotunneling ($t$), give an estimated temperature to observe the fractional plateau of $T\approx 170mK$, well within the experimental range of observability (see previous Section).}

\begin{acknowledgements}
DMP and RA acknowledge financial support from the Horizon Europe Framework Program of the European Commission through the European Innovation Council Pathfinder Grant No. 101115315 (QuKiT), the Spanish Ministry of Science through Grants No. PID2021-125343NB-I00 and No. TED2021-130292B-C43 funded by MCIN/AEI/10.13039/501100011033, “ERDF A way of
making Europe”, the European Union Next Generation
EU/PRTR and the State Research Agency through the predoctoral Grant No. PRE2022-103741 under the Program "State Program to Develop, Attract and Retain Talent", as well as the CSIC Interdisciplinary Thematic
Platform (PTI+) on Quantum Technologies (PTI-QTEP+). YM acknowledges support from the European
Research Council (ERC) under the European Unions
Horizon 2020 research and innovation programme under
Grant Agreement No. 951541 
and the Israel Science Foundation Grant No. 154/19.
\end{acknowledgements}

{\bf Note added} While finishing this manuscript, two recent preprints in Arxiv \cite{shen2024nonhermitian} and \cite{beenakker2024josephson} also pointed out the subtleties of calculating the free energy in a NH Josephson junction. Eq. (9) in Ref. \cite{shen2024nonhermitian} for the supercurrent agrees with our Eq. \eqref{supercurrent} in the limit $D\to\infty$. Moreover, Eq. (16) in Ref. \cite{beenakker2024josephson} agrees with our Eq. \eqref{supercurrent} in the regime without EPs. This latter case, in particular, results in a reduction factor $\frac{2}{\pi}\arctan\left(\frac{E}{\gamma}\right)$ in the $T\to 0$ supercurrent, see Eq. \eqref{eq:T-limit_supercurrent_ABS}, owing to the coupling with the reservoir.
\bibliography{bibliography}

\begin{widetext}

\appendix

\section{Partition function formalism}\label{sec:Appendix_A}
The partition function of a generic quantum system can be written as
\begin{equation} \label{eq:partition-function}
    Z = \Tr e^{-\beta(H - \mu N)} = \prod_n \left( 1 + e^{-\beta(\epsilon_n-\mu)}\right) \;.
\end{equation}
\editR{Here, $\mu$ is the chemical potential and $\beta=1/k_BT$, the inverse temperature. The second step explicitly assumes a diagonal form in terms of energies $\epsilon_n$. Using this partition function, the Helmholtz free energy of the system reads
\begin{equation}\label{eq:free-energy_def}
    F = -\frac{1}{\beta}\log Z=-\frac{1}{\beta}\sum_n \log \left( 1 + e^{-\beta(\epsilon_n-\mu)}\right)=-\frac{1}{\beta}\int_{-\infty}^\infty d\omega \, \rho(\omega) \log\left(1 + e^{-\beta(\omega-\mu)}\right) \;,
\end{equation}
where $\rho(\omega)\equiv\sum_n \delta(\omega-\epsilon_n)$ is the density of states of the system.
Differentiating the free energy expression with respect to the chemical potential on gets the average occupation
\begin{equation}
\label{FandN}
    -\frac{\partial F}{\partial\mu} = \int_{-\infty}^\infty d\omega \, \rho(\omega) f(\omega-\mu) =\int_{-\infty}^\infty d\omega G^< (\omega)=\langle N\rangle \;,
\end{equation}
with $f(\omega-\mu)=1/ \left( 1 + e^{\beta(\omega-\mu)}\right)$ being the Fermi-Dirac distribution function and $G^< (\omega)$ the so-called Keldysh lesser Green's function.
This expression connecting the occupation with the derivative of the free energy with respect to the chemical potential will be very useful later on.}

\editR{In what follows, we focus on a single pole $E-i\gamma$. In the models discussed in the main text, this describes how the eigenstates of the system acquire an imaginary part (corresponding to a finite lifetime)  when there is a finite coupling to an electron reservoir (within the so-called wideband limit \cite{PhysRevLett.68.2512}). This results in a density of states,}
\begin{equation}
    \rho(\omega)  =\frac{1}{\pi}\Im\partial_\omega \log G^R(\omega) = \frac{\gamma/\pi}{(\omega-E)^2 + \gamma^2} \;.
\end{equation}
Using this density of states, 
we can integrate by parts Eq. (\ref{eq:free-energy_def}) and obtain
\begin{equation} \label{eq:free-energy_Hewson}
    F = \left. \frac{1}{\pi}\log\left(1 + e^{-\beta(\omega-\mu)}\right)\arctan\frac{\gamma}{\omega - E}\right|_{-\infty}^\infty + \frac{1}{\pi} \int_{-\infty}^\infty d\omega \, f(\omega-\mu) \arctan\frac{\gamma}{\omega - E} \;,
\end{equation}
where we have expanded the complex logarithm of $G^R(\omega)$ as
\begin{equation}
    \log\left(\frac{1}{\omega - E + i\gamma}\right) = -\frac{1}{2}\log\left[ (\omega-E)^2 + \gamma^2 \right] - i\arctan\frac{\gamma}{\omega - E} \;.
\end{equation}

If we neglect the first term in (\ref{eq:free-energy_Hewson}), we \editR{recover Eq. (7.97) from Hewson's book \cite{Hewson} when} taking the wide flat band approximation $-D<\omega<D$. At $T=0$ this result is exact,
\begin{equation} \label{eq:free-energy_Hewson_T-0}
\begin{aligned}
    F_{T=0} & = \frac{1}{\pi}\int_{-D}^\mu d\omega \, \arctan\frac{\gamma}{\omega-E} = \frac{1}{\pi}\int_{-D}^\mu d\omega \, \mathrm{arccot}\frac{\omega-E}{\gamma} = \frac{\gamma}{\pi}\int_{-\frac{D+E}{\gamma}}^{\frac{\mu-E}{\gamma}} dz \, \mathrm{arccot}(z)
    \\
    & = \left. \frac{\gamma}{\pi} z\,\mathrm{arccot}(z) + \frac{\gamma}{2\pi}\log(1+z^2)\right|_{-\frac{D+E}{\gamma}}^{\frac{\mu-E}{\gamma}}
    \\
    & = \frac{\gamma}{\pi} - \frac{E-\mu}{\pi}\arctan\frac{E-\mu}{\gamma} - \frac{E-\mu}{2} + \frac{\gamma}{\pi}\log\frac{\sqrt{\gamma^2 + (\mu-E)^2}}{D} \;.
\end{aligned}
\end{equation}
We have taken the limit $D\to\infty$, since this parameter is the largest energy scale of the problem.

\section{Free energy from the occupation}\label{sec:Appendix_B}

\editR{While the zero-temperature result in the previous example can be found in many books, its generalization to finite $T$ through  Eq. (\ref{eq:free-energy_def}) is highly non-trivial, even for the simple model considered here, so we must search for an alternative way to calculate the free-energy from an analytically manageable expression.}

\editR{The difficulty of the calculation can be intuited just by considering the simplest case in which the problem has been diagonalized such that the full complex pole structure of the Green's function is known, $\epsilon_n=E_n-i\gamma_n$.  
Even in this simple case, a direct development from the partition function (\ref{eq:partition-function}) could give rise to a complex-valued free energy \cite{PhysRevB.106.L121102,cayao2023nonhermitian,li2024anomalous}. This situation is being treated in the literature in various, somewhat contradictory, ways that in our opinion need further investigation. In order to avoid possible inconsistencies in the calculation, we instead use Eq. (\ref{FandN}) to analytically calculate the free energy through the occupation. This occupation is in general a well-defined quantity in an open quantum system, even in non-equilibrium situations where $G^< (\omega)$ can be generalized beyond the fluctuation-dissipation theorem \cite{PhysRevLett.68.2512}.}

\subsection{Analytical expressions for the occupation and the free energy}

The first step is to use the following analytical integral,
\begin{equation}
\begin{aligned}
    & I = \int d\omega \, \frac{D^2}{(\omega+iD)(\omega-iD)}\frac{1}{\omega-E+i\gamma} f(\omega-\mu)
    = \frac{D^2}{(E-i\gamma+iD)(E-i\gamma-iD)} \psi\left(\frac{1}{2}-\frac{(E-i\gamma-\mu)\beta}{2i\pi}\right)
    \\
    & + \frac{D/(2i)}{E-i\gamma+iD}\psi\left(\frac{1}{2}+\frac{(iD+\mu)\beta}{2i\pi}\right)
    - \frac{D/(2i)}{E-i\gamma-iD}\psi\left(\frac{1}{2}+\frac{(iD-\mu)\beta}{2i\pi}\right) - \frac{\pi D/2}{E-i\gamma-iD} \;,
\end{aligned}
\end{equation}
where we have included a Lorentzian cutoff $\Omega(\omega,D)=D^2/(\omega^2+D^2)$, which will be relevant later. Using this integral, we can write the occupation as $\langle N\rangle=-1/\pi\Im(I)$, which gives
\begin{equation} \label{eq:occupation}
\begin{aligned}
    \langle N\rangle & = \frac{-D^2}{\pi}\frac{(E^2-\gamma^2+D^2)\Im\tilde{\psi}(i\gamma-E+\mu) + 2\gamma E\Re\tilde{\psi}(i\gamma-E+\mu)}{[E^2+(D-\gamma)^2][E^2+(D+\gamma)^2]} + \frac{D}{2}\frac{D+\gamma}{E^2+(D+\gamma)^2}
    \\
    & + \frac{D}{2\pi}\frac{(D-\gamma)\Im\tilde{\psi}(iD+\mu) + E\Re\tilde{\psi}(iD+\mu)}{E^2+(D-\gamma)^2} + \frac{D}{2\pi}\frac{(D+\gamma)\Im\tilde{\psi}(iD-\mu) - E\Re\tilde{\psi}(iD-\mu)}{E^2+(D+\gamma)^2} \;,
\end{aligned}
\end{equation}
where $\tilde{\psi}(z) = \psi[1/2 + z/(2i\pi T)]$ are digamma functions. In the $D\to\infty$, the occupation becomes
\begin{equation} \label{eq:occupation_approx}
    \langle N\rangle \approx -\frac{1}{\pi}\Im\psi\left(\frac{1}{2}+\frac{i\gamma-E+\mu}{2i\pi T}\right) + \frac{1}{2} \;,
\end{equation}
which coincides with the Eq. (16.249) from Coleman's book \cite{Coleman}
\begin{equation}
    I_2 = \int d\omega \frac{D^2}{D^2+\omega^2}\frac{f(\omega)}{\omega-\xi} = \psi\left(\frac{1}{2}+\frac{\xi}{2\pi iT}\right) - \log\frac{D}{2\pi T} + i\frac{\pi}{2}
\end{equation}
from where we can calculate the occupation as $\langle N\rangle= +1/\pi \Im(I_2)$, since this approach has been done with advanced poles ($\xi=E+i\gamma$). It has also been used $\psi(z)\to\log(z)$ for large $z$.

\begin{figure}[ht]
    \centering
    \includegraphics[width=0.8\linewidth]{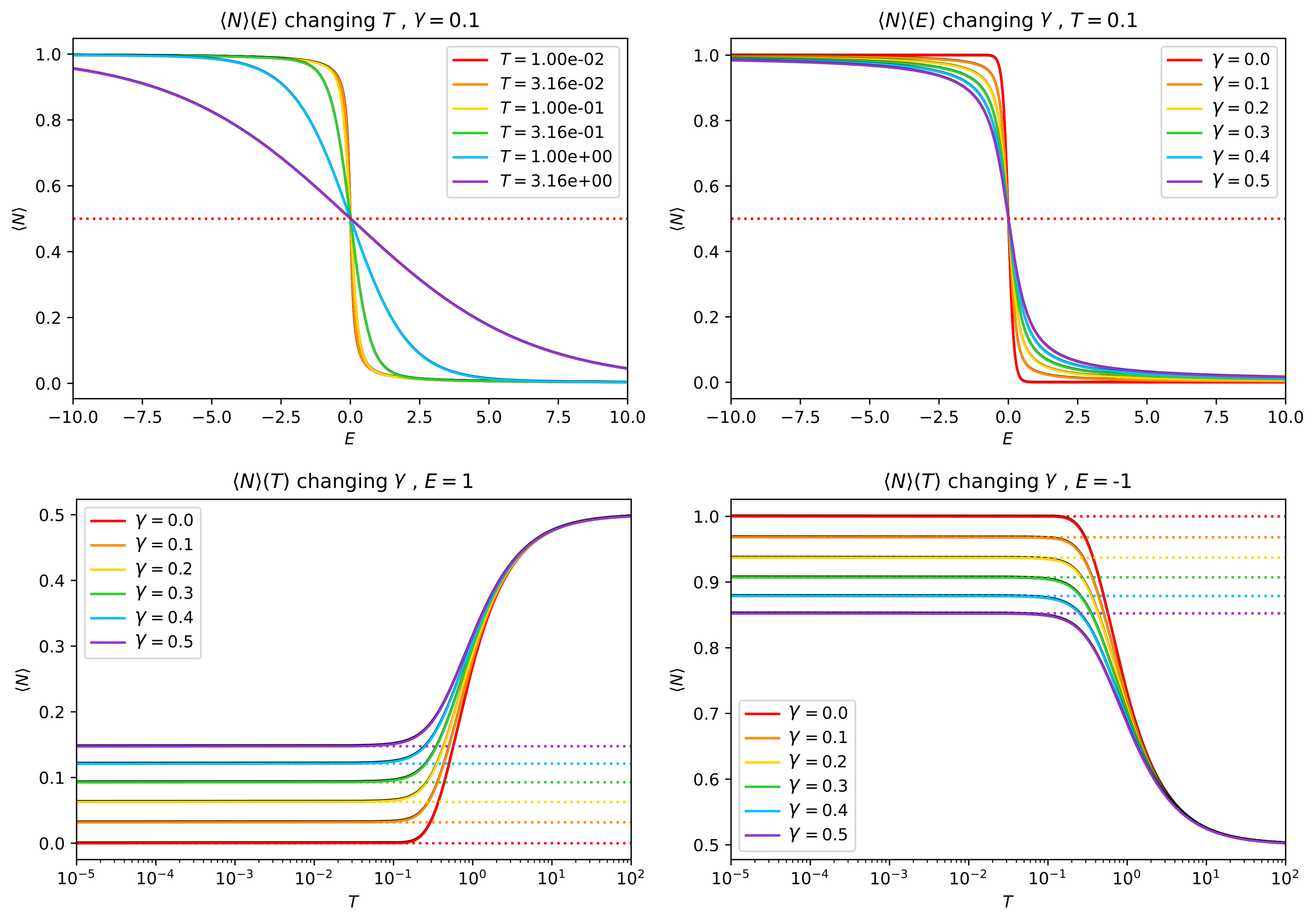}
    \caption{\textbf{Evolution of occupation} as a function of $E$ and $T$. Colored lines correspond to the full expression (\ref{eq:occupation}), whereas black lines refer to the approximated result (\ref{eq:occupation_approx}). Colored dotted lines indicate the $T\to 0$ limit (\ref{eq:occupation_T-0}). The cutoff parameter $D$ is always chosen to be the major quantity of the system, so, unless otherwise stated, $D=100$.}
    \label{fig:occupation}
\end{figure}

Top panel of Fig. \ref{fig:occupation} shows the occupation as a function of $E$, changing $T$ and $\gamma$, respectively. As it is expected, our expression recovers the asymptotic behavior $\langle N\rangle\to 1$ when $E\to-\infty$ and $\langle N\rangle\to 0$ when $E\to\infty$. The effect of both $T$ and $\gamma$ is a broadening of the jump around $E=0$, where $\langle N\rangle=1/2$ always. 

On the other hand, the behavior of the occupation when evolving $T$ is also well recovered (bottom panel of Fig. \ref{fig:occupation}). Indeed, when $T\gg 1$, $\langle N\rangle\to 1/2$, and for small temperatures, it approaches the analytical value
\begin{equation} \label{eq:occupation_T-0}
    \langle N\rangle(T=0) = \frac{1}{2} - \frac{1}{\pi}\arctan\frac{E}{\gamma} \;,
\end{equation}
in agreement with Eq. (5.47) from Hewson's book \cite{Hewson}. All of these previous results are plotted using the full Eq. (\ref{eq:occupation}) --colored lines-- and they are also compared with the approximated expression (\ref{eq:occupation_approx}) --black lines--, leading to a good agreement between them.

Once we know the occupation (\ref{eq:occupation}), it is straightforward to calculate the exact free energy as
\begin{equation}
F=-\int d\mu\langle N\rangle=\frac{1}{\pi}\int d\mu\Im(I) \;,
\end{equation}
resulting in
\begin{equation} \label{eq:SI_free-energy}
\begin{aligned}
    F & = 2TD^2\frac{(E^2-\gamma^2+D^2)\Re\log\tilde{\Gamma}(i\gamma-E+\mu) - 2\gamma E\Im\log\tilde{\Gamma}(i\gamma-E+\mu)}{[E^2+(D-\gamma)^2][E^2+(D+\gamma)^2]} - \frac{D}{2}\frac{\mu(D+\gamma)}{E^2+(D+\gamma)^2} + h(T,E,\gamma)
    \\
    & - TD\frac{(D-\gamma)\Re\log\tilde{\Gamma}(iD+\mu) - E\Im\log\tilde{\Gamma}(iD+\mu)}{E^2+(D-\gamma)^2} + TD\frac{(D+\gamma)\Re\log\tilde{\Gamma}(iD-\mu) \michel{+} E\Im\log\tilde{\Gamma}(iD-\mu)}{E^2+(D+\gamma)^2} \;,
\end{aligned}
\end{equation}
where $h(T,E,\gamma)$ is a generic function that comes from the integration ``constant'' concerning $\mu$. Taking the limit $D\to \infty$, we can approximate the free energy (\ref{eq:SI_free-energy}) as
\michel{\begin{equation}\label{eq:free-energy_D-limit}
    F_{D\to\infty} = 2T\Re\log\Gamma\left(\frac{1}{2} + \frac{i\gamma-E+\mu}{2i\pi T}\right) - \frac{2T\gamma}{D}\Re\log\Gamma\left(\frac{1}{2} + \frac{iD+\mu}{2i\pi T}\right) - \frac{\mu}{2} + h(T,E,\gamma) \;.
\end{equation}}
\editR{Alternatively,} we can also directly extract an approximated expression for the free energy from the occupation in Eq. (\ref{eq:occupation_approx}),
\begin{equation} \label{eq:free-energy_approx}
\begin{aligned}
    & F \approx 2T\Re\log\Gamma\left(\frac{1}{2}+\frac{i\gamma-E+\mu}{2i\pi T}\right) - \frac{\mu}{2} + h(T,E,\gamma) \;.
\end{aligned}
\end{equation}

This discrepancy between both approximations will be relevant for avoiding divergences in the low-temperature asymptotic behavior of the free energy and also the entropy.
\begin{figure}[ht]
    \centering
    \includegraphics[width=0.8\linewidth]{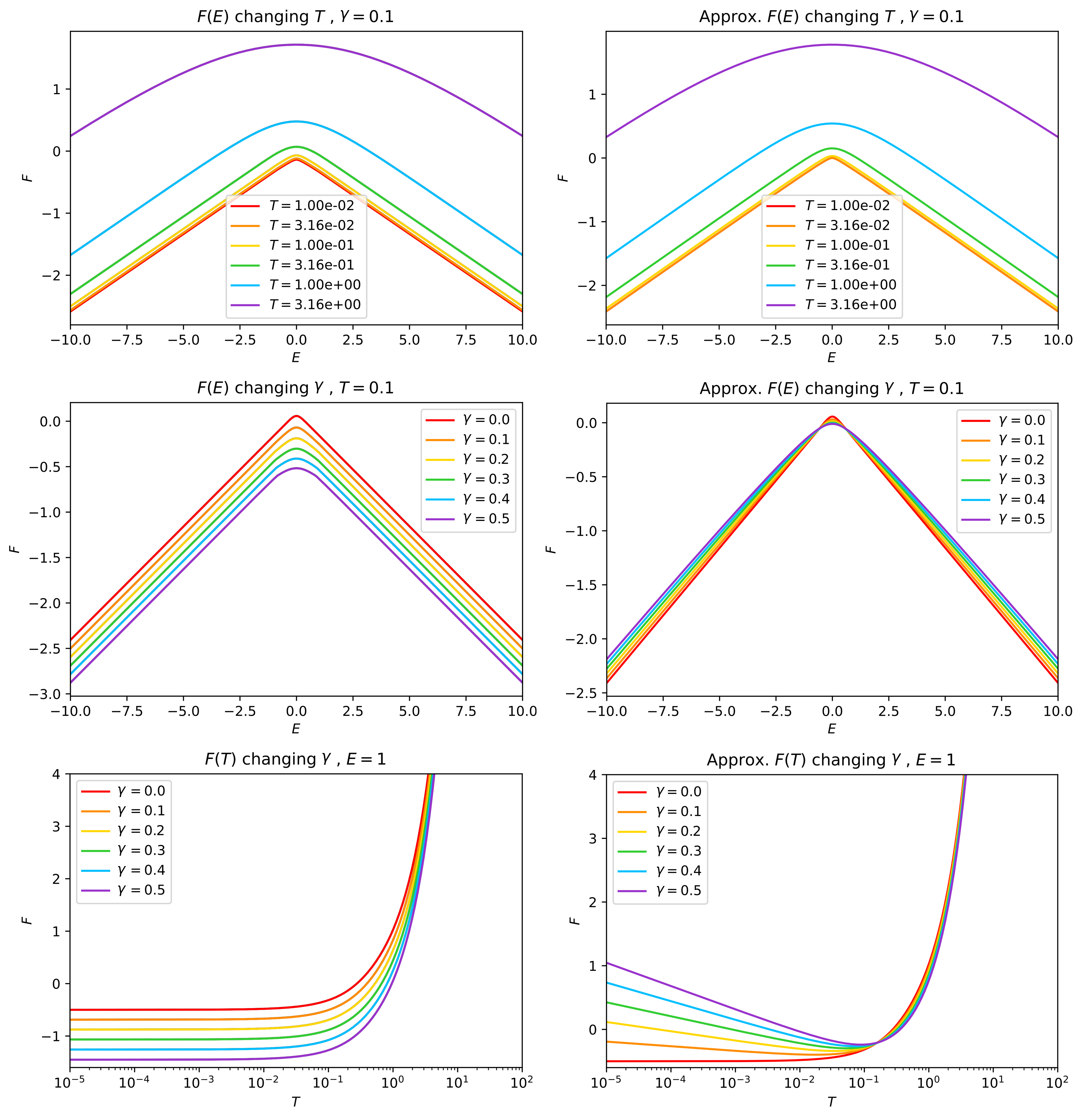}
    \caption{\textbf{Evolution of free energy} as a function of $E$ and $T$. \textbf{Left:} (colored lines) full expression (\ref{eq:SI_free-energy}) and (black lines) its $D\to\infty$ limit (\ref{eq:free-energy_D-limit}). \textbf{Right:} approximated expression (\ref{eq:free-energy_approx}).}
    \label{fig:free-energy}
\end{figure}

\subsection{Why are cutoffs important?}

In what follows we elaborate on the importance of keeping a finite cutoff in the free energy calculations (and later on their relevance for the low-temperature asymptotics of the entropy). \editR{While naively one would expect that all the approaches give reasonable results, since the occupations, c.f Fig. \ref{fig:occupation}, are properly captured, even for the seemingly simple single pole, full (\ref{eq:free-energy_D-limit}) and approximated (\ref{eq:free-energy_approx}) expressions for the free energy are no longer equal when $\gamma$ is non-negligible. This is illustrated in Fig. \ref{fig:free-energy}. Most importantly, this disagreement will translate to divergent behavior of the entropy, as we will see later}. 

This discrepancy between results can be \editR{understood} from the asymptotic behavior of the free energy for low temperatures. Since the log-gamma function grows as
\begin{equation}
\begin{aligned}
    & \log\Gamma(z) = z(\log(z) - 1) - \frac{1}{2}\log(z) + \frac{1}{2}\log(2\pi) + \mathcal{O}(1/z)
    \\
    & \Re\log\Gamma(z) = x(\log|z| - 1) - y\arg(z) - \frac{1}{2}\log|z| + \frac{1}{2}\log(2\pi) + \mathcal{O}(1/z) \qquad (z=x+iy)
    \\
    & \Im\log\Gamma(z) = x\arg(z) + y(\log|z|-1) - \frac{1}{2}\arg(z) + \mathcal{O}(1/z)
\end{aligned}
\end{equation}
when $|z|\to\infty$ at constant $\arg(z)<\pi$, from (\ref{eq:free-energy_D-limit}) the $T\to 0$ limit of the free energy takes a very compact form,
\michel{\begin{equation} \label{eq:free-energy_Tlimit}
\begin{aligned}
    F_{T\to0} = \frac{\gamma}{\pi}\log\left(\frac{\sqrt{\gamma^2+(E-\mu)^2}}{D}\right) - \frac{E-\mu}{\pi}\arctan\frac{E-\mu}{\gamma} - \frac{\mu}{2} + h(0,E,\gamma)
\end{aligned}
\end{equation}}
while, from (\ref{eq:free-energy_approx}), we have
\begin{equation}
\begin{aligned}
    F_{T\to 0} = \frac{\gamma}{\pi}\left[ \log\left(\frac{\sqrt{\gamma^2+(E-\mu)^2}}{2\pi T}\right) - 1 \right] - \frac{E-\mu}{\pi}\arctan\frac{E-\mu}{\gamma} - \frac{\mu}{2} + h(0,E,\gamma)
\end{aligned}
\end{equation}

As one can see, the inclusion of the cutoff leads to an $T$-independent asymptotic behavior of the free energy, whereas the ``naked'' free energy has a Kondo-like divergence $\sim \log(1/T)$ in this regime when $\gamma$ is finite. In contrast, the limit $\gamma\to 0$ is well defined for both expressions (\ref{eq:free-energy_D-limit}) and (\ref{eq:free-energy_approx}),
\begin{equation}
\begin{aligned}
    F_{\gamma=0} =& T\log(2\pi) - T\log\left(2\cosh\frac{E-\mu}{2T}\right) - \frac{\mu}{2} + h(T,E,0)
\end{aligned}
\end{equation}
Indeed, the weak coupling regime leads to an agreement between full and approximated free energy (top panels of Fig. \ref{fig:free-energy}). However, in the strong coupling limit, both approaches differ (center panels of Fig. \ref{fig:free-energy}), and the inclusion of a cutoff prevents divergences at low temperatures (bottom panels of Fig. \ref{fig:free-energy}). Comparing both limits $T\to0$ and $\gamma\to0$ with the textbook results
\begin{equation}\label{eq:textbook-limits}
\begin{aligned}
    F_{T\to 0} &= -\frac{\gamma}{\pi} + \frac{E-\mu}{2} - \frac{E-\mu}{\pi}\arctan\frac{E-\mu}{\gamma} + \frac{\gamma}{\pi}\log\left(\frac{\sqrt{\gamma^2+(E-\mu)^2}}{D}\right)
    \\
    F_{\gamma=0} &= \frac{E-\mu}{2} - T\log\left(2\cosh\frac{E-\mu}{2T}\right)
\end{aligned}
\end{equation}
we can fix the generic function $h(T,E,\gamma)=-\gamma/\pi-T\log(2\pi)$. Note that we have not included the term $E/2$ that appears in these well-known limits since they cancel when our formalism is extended to BdG Hamiltonians. Indeed, these expressions are written for a single particle state $\epsilon^p=E-i\gamma$ with positive real part, $E>0$. On the contrary, for its hole-branch counterpart $\epsilon^h=-E-i\gamma$, expressions (\ref{eq:textbook-limits}) are transformed such that $E-\mu\to-E-\mu$. This change of sign cancels the term $E/2$ when considering both particle and hole states.

\section{Extension to BdG Hamiltonians}\label{sec:Appendix_C}

The previous developments done for a single pole can be easily extended to a BdG Hamiltonian, which, for each particle state $\epsilon^p=E-i\gamma$, integrates its hole-branch counterpart $\epsilon^h=-E-i\gamma$, related by particle-hole symmetry $\epsilon^h=-(\epsilon^p)^*$. Hence, the total spectral function
\begin{equation}
    A(\omega) = -\frac{1}{2\pi}\Im\sum_\epsilon \frac{1}{\omega - \epsilon} = \frac{1}{2}\left[\rho^p(\omega) + \rho^h(\omega)\right]
\end{equation}
takes into account both particle and hole branches, where a global factor $1/2$ must be added due to the duplicity of dimensions that emanates from BdG formalism. At the same time, $\rho^{p/h}(\omega)$ is a sum over all the $p/h$-states. All the previous quantities can be calculated for the hole branch by making the substitution $f(\omega)\to f(-\omega)$, such that
\begin{equation}
    \langle N\rangle = \int d\omega \, A\cdot f = \frac{1}{2}\int d\omega \, \rho^p(\omega) f(\omega-\mu) + \frac{1}{2}\int d\omega \, \rho^h(\omega) f(-\omega-\mu)
\end{equation}
and similarly for the free energy and entropy. This treatment leads to the same analytical results (\ref{eq:occupation},\ref{eq:SI_free-energy}) that we have obtained for the particle branch. Indeed, the hole-branch integral can be seen as a reflection on the real axis of the particle-branch one, so that both of them lead to the same result. \michel{For this reason, we always restrict $E,\gamma>0$, since these quantities are those who appear in the analytical results: although $\Re(\epsilon^h)<0$, every quantity is defined in terms of $E=\Re(\epsilon^p)=-\Re(\epsilon^h)$, so that, for a generic pole $\epsilon$, $E$ will be generally defined as $|\Re(\epsilon)|$ whereas the imaginary part is always retarded, $\gamma=|\Im(\epsilon)|=-\Im(\epsilon)$.}
Thus, both particle and hole poles contribute the same to every quantity ($O^p=O^h$), since the magnitude of their real and imaginary parts are identical. This is a consistent result with BdG formalism, since in this framework particles and holes are two copies referring to the same excitation. Hence, the total thermodynamic quantities can be written as a sum over all the states, $O_{\mathrm{tot}} = \frac{1}{2}\sum_j O_j$, where $O$ can be $\langle N\rangle$ or $F$.

A case of special interest is the appearance of exceptional points (EPs) in the spectrum, where the particle and hole branches of the ground state coalesce, $\epsilon^p_g=\epsilon^h_g=-i\gamma$. After the EP, both eigenvalues bifurcate, acquiring different (full imaginary) values, $\epsilon^+=-i\gamma^+$ and $\epsilon^-=-i\gamma^-$ such that $\gamma_\mathrm{asym}\equiv\frac{\gamma^- - \gamma^+}{2}\neq 0$. Although these two pure imaginary --but distinct-- eigenvalues keep particle-hole symmetry, this difference of magnitude implies $O^+\neq O^-$. Hence, each state will draw a particular characteristic curve in occupation and free energy. 
In what follows we discuss the relevance of EPs when calculating the Josephson current of a non-Hermitian junction.

\subsection{Non-Hermitian Josephson junctions: role of EPS}
One relevant quantity that can be extracted from the free energy is the supercurrent of a non-hermitian Josephson junction. For a general pair of BdG poles $\epsilon^\pm(\phi)=\pm E(\phi)-i\gamma(\phi)$, the supercurrent across the junction is defined from (\ref{eq:free-energy_D-limit}) as
\begin{equation}
    I = \frac{\partial F}{\partial \phi} = -\frac{1}{\pi}\Im\psi\left(\frac{1}{2}+\frac{i\gamma-E}{2i\pi T}\right)\frac{\partial E}{\partial\phi} + \frac{1}{\pi}\Re\psi\left(\frac{1}{2}+\frac{i\gamma-E}{2i\pi T}\right)\frac{\partial \gamma}{\partial\phi} - \frac{2T}{D}\log\Gamma\left(\frac{1}{2}+\frac{D}{2\pi T}\right)\frac{\partial\gamma}{\partial\phi} - \frac{1}{\pi}\frac{\partial\gamma}{\partial\phi}
\end{equation}
since both particle and hole poles contribute the same to the supercurrent. Conversely, after an EP, this is not longer true, and the supercurrent becomes a sum $I = \frac{1}{2}\sum_\pm I^\pm$ over two independent states $\epsilon^\pm(\phi) = -i\gamma^\pm(\phi)$. \michel{Note that since $E=|\Re(\epsilon)|$, its derivative will be $\partial_\phi E = \sgn[\Re(\epsilon)]\Re(\partial_\phi\epsilon)$, whereas $\partial_\phi\gamma=-\Im(\partial_\phi\epsilon)$.} In the low-temperature regime, the supercurrent can be written as
\begin{equation} \label{eq:T-limit_supercurrent}
\begin{aligned}
    \text{before EP:} \quad & I_{T\to0} = -\frac{1}{\pi}\arctan\left(\frac{E}{\gamma}\right)\frac{\partial E}{\partial\phi} + \frac{1}{\pi}\log\left(\frac{\sqrt{\gamma^2+E^2}}{D}\right)\frac{\partial\gamma}{\partial\phi} - \frac{1}{\pi}\frac{\partial\gamma}{\partial\phi}
    \\
    \text{after EP:} \quad & I_{T\to0} = \frac{1}{2\pi}\log\left(\frac{\gamma^+}{D}\right)\frac{\partial\gamma^+}{\partial\phi} + \frac{1}{2\pi}\log\left(\frac{\gamma^-}{D}\right)\frac{\partial\gamma^-}{\partial\phi} - \frac{1}{2\pi}\left(\frac{\partial\gamma^+}{\partial\phi} + \frac{\partial\gamma^-}{\partial\phi}\right)
\end{aligned}
\end{equation}
where we have done $\psi(z)\approx \log(z) - 1/(2z)$ when $|z|\to\infty$ at constant $\arg(z)<\pi$. This asymptotic limit is independent of $T$ but it depends on the cutoff parameter $D$. However, we can demonstrate that this dependence is canceled by the symmetry of the BdG problem (see section III.D).

\subsection{ABS model}

Let's apply these results to a non-Hermitian Josephson junction, whose Hamiltonian is
\begin{equation}
    H^{JJ} = \left(\begin{matrix}
    -i\gamma_0^L & -i\Delta\sqrt{1-\tau\sin^2(\phi/2)} \\
    i\Delta\sqrt{1-\tau\sin^2(\phi/2)} & -i\gamma_0^R
    \end{matrix}\right)
\end{equation}
and describes a pair of ABS with complex energy
\begin{equation} \label{eq:ABS_eigenvalues}
    \epsilon^\pm = - \frac{i}{2}(\gamma_0^L+\gamma_0^R) \pm \frac{1}{2}\sqrt{2\Delta^2[2 + \tau(\cos\phi-1)] - (\gamma_0^L-\gamma_0^R)^2}
\end{equation}
which can present EPs only when $\delta_0=\gamma_0^L-\gamma_0^R\neq 0$, that is, when the discriminant can be negative,
\begin{equation} \label{eq:ABS_EP}
    \cos\phi < \frac{(\gamma_0^L-\gamma_0^R)^2 - 2\Delta^2(2-\tau)}{2\Delta^2\tau}
\end{equation}
Note that, unlike the Hermitian analog, there can appear zero energy crossings (EPs) even when $\tau<1$, Fig. \ref{fig:ABS_spectrum}. From the derivative of (\ref{eq:ABS_eigenvalues}), we can study separately the two phases of the spectrum,
\begin{equation}
\begin{aligned}
    \partial_\phi\epsilon^\pm &= \frac{\mp\Delta^2 \tau \sin\phi}{2\sqrt{{2\Delta^2[2+\tau(\cos\phi-1)] - (\gamma_0^L-\gamma_0^R)^2}}}
    \\
    \\
    \text{before EP} & \left\{ \begin{aligned}
        & \partial_\phi E = \frac{-\Delta^2 \tau \sin\phi}{2\sqrt{{2\Delta^2[2+\tau(\cos\phi-1)] - (\gamma_0^L-\gamma_0^R)^2}}}
        \\
        & \partial_\phi \gamma = 0
    \end{aligned} \right.
    \\
    \\
    \text{after EP} & \left\{ \begin{aligned}
        & \partial_\phi E = 0
        \\
        & \partial_\phi \gamma^\pm = \frac{\mp\Delta^2 \tau \sin\phi}{2\sqrt{{(\gamma_0^L-\gamma_0^R)^2 - 2\Delta^2[2+\tau(\cos\phi-1)]}}}
    \end{aligned} \right.
\end{aligned}
\end{equation}
we can extract the supercurrent before and after the EP,
\begin{equation}
\begin{aligned}
    \text{before EP:} \quad I &= -\frac{1}{\pi}\Im\psi\left(\frac{1}{2}+\frac{i\gamma-E}{2i\pi T}\right)\frac{\partial E}{\partial\phi}
    \\
    \\
    \text{after EP:} \quad I &= \frac{1}{2\pi}\sum_{\alpha=\pm}\left[\Re\psi\left(\frac{1}{2}+\frac{\gamma^\alpha}{2\pi T}\right) - 1\right]\frac{\partial \gamma^\alpha}{\partial\phi} = \pm\frac{1}{2\pi}\left[\Re\psi\left(\frac{1}{2}+\frac{\gamma^-}{2\pi T}\right)-\Re\psi\left(\frac{1}{2}+\frac{\gamma^+}{2\pi T}\right)\right]\frac{\partial \gamma^\pm}{\partial\phi}
\end{aligned}
\end{equation}

\begin{figure}[ht]
    \centering
    \includegraphics[width=0.8\linewidth]{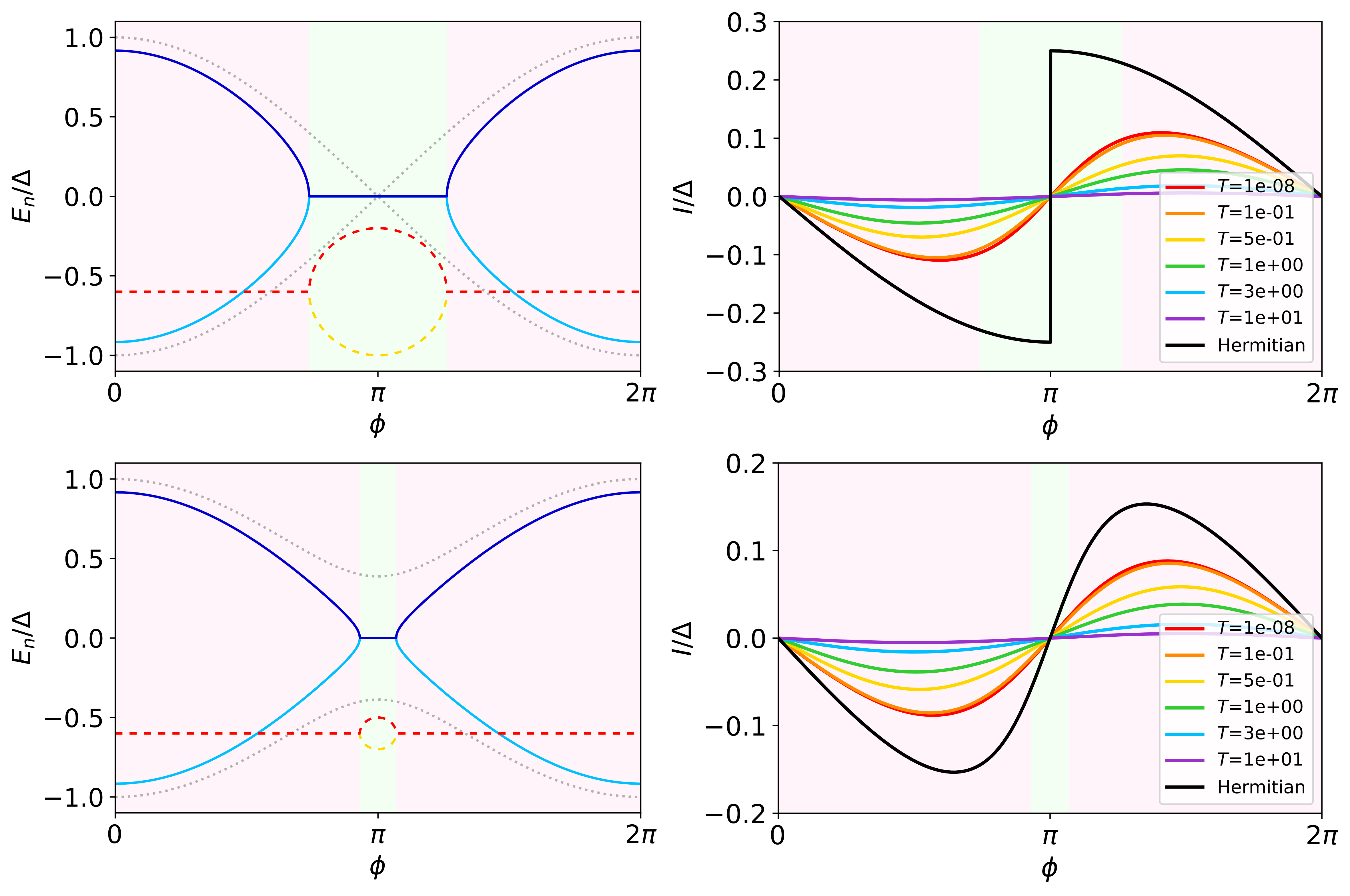}
    \caption{\textbf{ABS model}. \textbf{Left:} Energy spectrum of a pair of non-Hermitian ABS levels. Gray dotted lines correspond to the closed system ($\delta=0$). Lightblue/yellow solid lines correspond to real/imaginary parts of $\epsilon^-$, whereas darkblue/red dashed ones do to $\epsilon^+$. \textbf{Right:} Supercurrent as a function of $\phi$ for different temperatures. Green dotted lines mark the position of the EPs (\ref{eq:ABS_EP}). Black curves correspond to the Hermitian analog at $T=10^{-8}$. System parameters are fixed as $\Delta=1$ and $\gamma^L_0=5\gamma^R_0=1$. $\tau=1, 0.85$ in top and bottom panels, respectively.}
    \label{fig:ABS_spectrum}
\end{figure}

This quantity is also well-defined in the low-temperature regime, leading to an asymptotic behavior
\begin{equation} \label{eq:SI_T-limit_supercurrent_ABS}
\begin{aligned}
    \text{before EP:} \quad I_{T\to 0} & = \frac{1}{\pi}\arctan\left(\frac{\gamma}{E}\right) \frac{\partial E(\phi)}{\partial\phi}
    \\
    \text{after EP:} \quad I_{T\to 0} & = \frac{1}{2\pi}\log\left(\frac{\gamma^\pm}{\gamma^\mp}\right)\frac{\partial\gamma^\pm}{\partial\phi}
\end{aligned}
\end{equation}
In any case, both the free energy and supercurrent are free of divergences and/or do not contain any imaginary component, in contrast to previous approaches which just take the derivative of complex eigenvalues \cite{cayao2023nonhermitian,li2024anomalous}.

\subsection{Four Majorana model}

Another system of interest is a Josephson junction model with four Majoranas. This model is a low-energy projection of a Josophson junction between two DQDs \cite{Pino}, and its Hamiltonian is
\begin{equation}
   H^{4M} = i(\Delta_{12}-t_{12})\eta_1\eta_2 - it_{23}\cos(\phi/2)\eta_2\eta_3 + i(\Delta_{34}-t_{34})\eta_3\eta_4
\end{equation}
where we have set $\mu_i=0$ $\forall i$. For simplicity, we will assume both chains are equal ($\Delta_{12}=\Delta_{34}=\Delta$ and $t_{12}=t_{34}=t$). In the Majorana Nambu basis $\psi^\eta=(\eta_1,\eta_2,\eta_3,\eta_4)$, its BdG Hamiltonian takes the form
\begin{equation}
   H_{BdG}^{4M} = i\left(\begin{matrix}
       -\gamma^0_1 & \Delta-t & 0 & 0 \\
       t-\Delta & -\gamma^0_2 & t_{23}\cos(\phi/2) & 0 \\
       0 & -t_{23}\cos(\phi/2) & -\gamma^0_3 & \Delta-t \\
       0 & 0 & t-\Delta & -\gamma^0_4
   \end{matrix}\right)
\end{equation}
where we have also added different non-Hermitian terms $\gamma^0_i$ to each mode. When $t=\Delta$ (no intra-chain coupling), this system has the following eigenvalues,
\begin{equation}
\begin{aligned}
   \epsilon_\mathrm{outer}^\pm &= -i\gamma^0_{1/4}
   \\
   \epsilon_\mathrm{inner}^\pm &= -\frac{i}{2}(\gamma^0_2 + \gamma^0_3) \pm \frac{1}{2}\sqrt{2t_{23}^2 (1+\cos\phi) - (\gamma^0_2-\gamma^0_3)^2}
\end{aligned}
\end{equation}
and their derivatives before and after the EP,
\begin{equation}
\begin{aligned}
    \partial_\phi\epsilon_\mathrm{outer}^\pm=0 \quad &,\quad \partial_\phi\epsilon_\mathrm{inner}^\pm = \frac{\mp t_{23}^2 \sin\phi}{2\sqrt{2t_{23}^2(1+\cos\phi) - (\gamma^0_2-\gamma^0_3)^2}}
    \\
    \\
    \text{before EP} & \left\{ \begin{aligned}
        & \partial_\phi E_\mathrm{inner} = \frac{-t_{23}^2 \sin\phi}{2\sqrt{2t_{23}^2(1+\cos\phi) - (\gamma^0_2-\gamma^0_3)^2}}
        \\
        & \partial_\phi \gamma_\mathrm{inner} = 0
    \end{aligned} \right.
    \\
    \\
    \text{after EP} & \left\{ \begin{aligned}
        & \partial_\phi E_\mathrm{inner} = 0
        \\
        & \partial_\phi \gamma^\pm_\mathrm{inner} = \frac{\mp t_{23}^2 \sin\phi}{2\sqrt{(\gamma^0_2-\gamma^0_3)^2 - 2t_{23}^2(1+\cos\phi)}}
    \end{aligned} \right.
\end{aligned}
\end{equation}
Note that the outermost states correspond to Majorana zero modes (MZMs) localized at the edges and they do not contribute to the supercurrent since they do not disperse with $\phi$. This case is similar to the ABS system with $\tau=1$, but in this case, the zero energy crossing is protected by the presence of MZMs and thus it does not depend on the transparency of the junction. Indeed, when $\gamma^0_2\neq\gamma^0_3$, there appear EPs around $\phi=\pi$, whereas if these couplings are equal ($\gamma^0_2=\gamma^0_3=\gamma_0$), the model defines a perfect 4$\pi$-Josephson effect with finite reservoir couplings: $\epsilon_\mathrm{inner}^\pm = -i\gamma_0 \pm t_{23}\cos(\phi/2)$.

Fig. \ref{fig:4Majorana-decoupled_spectrum} shows the spectrum of this configuration and its relative supercurrent, which is also well-defined in the low-temperature regime, showing an asymptotic behavior similar to (\ref{eq:SI_T-limit_supercurrent_ABS}).

\begin{figure}[ht]
    \centering
    \includegraphics[width=0.8\linewidth]{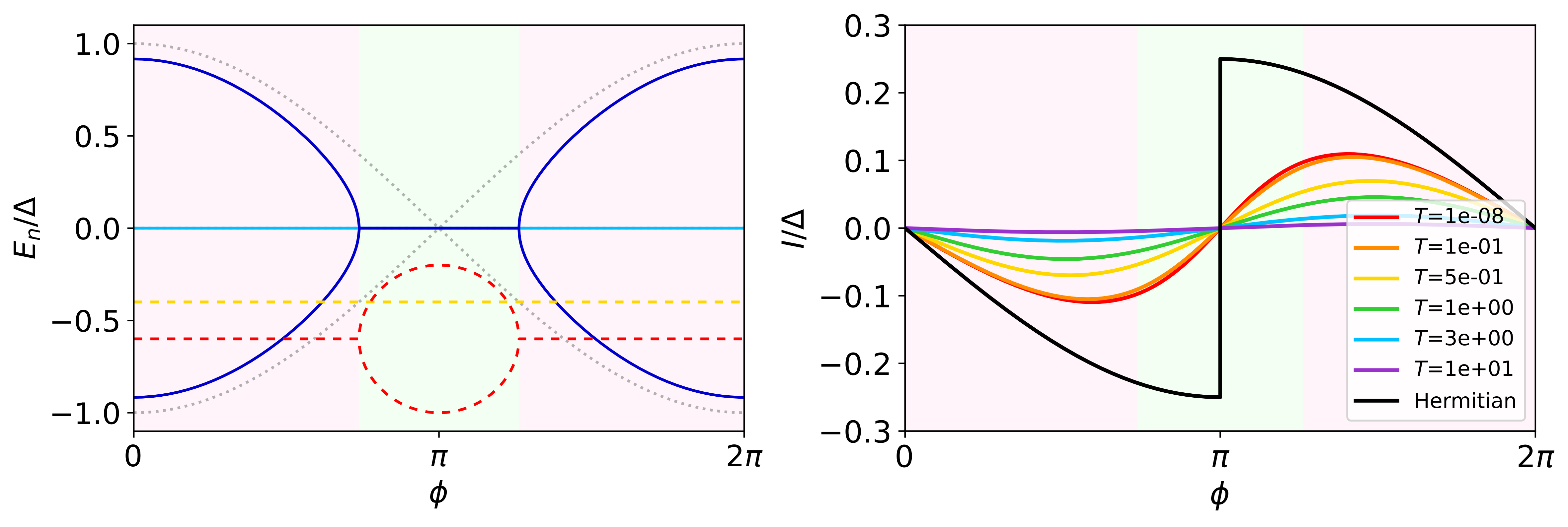}
    \caption{\textbf{Four Majorana model without intra-chain coupling}. \textbf{Left:} Energy spectrum of the model. Gray dotted lines correspond to the closed system ($\gamma^0_i=0$). Lightblue/yellow lines correspond to real/imaginary parts of $\epsilon^\pm_\mathrm{outer}$, whereas darkblue/red ones do to $\epsilon^\pm_\mathrm{inner}$. Dashed/solid lines distinguish $+/-$ states. \textbf{Right:} Supercurrent as a function of $\phi$ for different temperatures. The black curve corresponds to the Hermitian analog at $T=10^{-8}$. Green dotted lines mark the position of the EPs in both panels. System parameters are fixed as $t=\Delta=1$, $t_{23}=\Delta$, $\gamma^0_2=5\gamma^0_3=1$ and $\gamma^0_1=\gamma^0_4=0.4$.}
    \label{fig:4Majorana-decoupled_spectrum}
\end{figure}

On the other hand, turning on the intra-chain couplings ($\Delta-t\neq0$) gives the following eigenvalues,
\begin{equation}
\begin{aligned}
    \epsilon_A^\pm &= -\frac{i}{4}\gamma_0 - \frac{1}{4}\sqrt{\Lambda(\phi) - \delta_0^2} \pm \frac{1}{4}\sqrt{16(\Delta-t)^2 + \left(\sqrt{\Lambda(\phi)-\delta_0^2} + i\gamma_0\right)^2}
    \\
    \epsilon_B^\pm &= -\frac{i}{4}\gamma_0 + \frac{1}{4}\sqrt{\Lambda(\phi) - \delta_0^2} \pm \frac{1}{4}\sqrt{16(\Delta-t)^2 + \left(\sqrt{\Lambda(\phi)-\delta_0^2} - i\gamma_0\right)^2}
    \\
    \\
    &\partial_\phi\epsilon_A^\pm = \frac{\pm\epsilon_A^\pm t_{23}^2\sin\phi}{\sqrt{\Lambda(\phi)-\delta_0^2}\sqrt{16(\Delta-t)^2+\left(\sqrt{\Lambda(\phi)-\delta_0^2}+i\gamma_0\right)^2}}
    \\
    &\partial_\phi\epsilon_B^\pm = \frac{\mp\epsilon_B^\pm t_{23}^2\sin\phi}{\sqrt{\Lambda(\phi)-\delta_0^2}\sqrt{16(\Delta-t)^2+\left(\sqrt{\Lambda(\phi)-\delta_0^2}-i\gamma_0\right)^2}}
\end{aligned}
\end{equation}
where $\Lambda(\phi)=2t_{23}^2(1+\cos\phi)$, $\gamma^0_1=\gamma^0_4=0$, $\gamma_0=\gamma^0_2+\gamma^0_3$ and $\delta_0=\gamma^0_2-\gamma^0_3$. Fig. \ref{fig:4Majorana-twochains_spectrum} shows the spectrum and supercurrent for this configuration. We can see that for some particular parameter choice, two pairs of EPs appear, dividing the spectrum into three different phases. We also can see a slight enhancement of the maximum supercurrent $I_\mathrm{max}$ for some finite temperature that depends on the intra-chain coupling $\Delta-t$, even in the Hermitian analog. This anomalous behavior is displayed in Fig. \ref{fig:I-max}, where an enhancement of $I_\mathrm{max}$ appears at some ``critical'' temperature that increases with $\Delta-t$.

\begin{figure}[ht]
    \centering
    \includegraphics[width=0.8\linewidth]{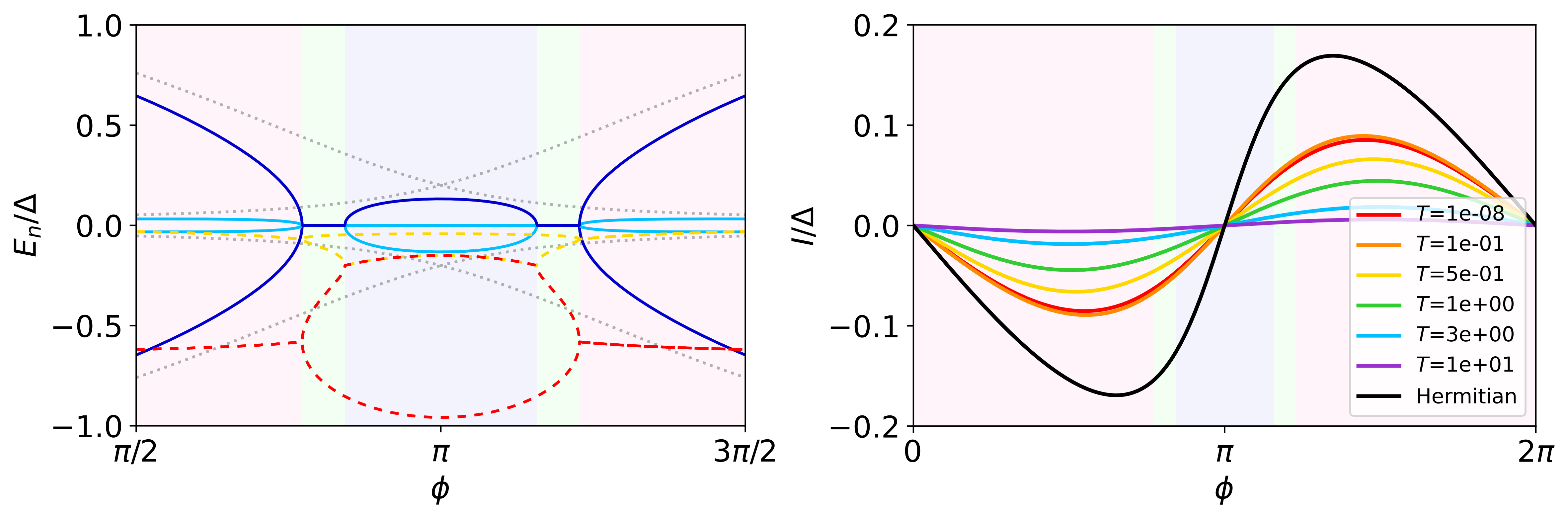}
    \caption{\textbf{Four Majorana model with intra-chain coupling}. \textbf{Left:} Energy spectrum of the model. Gray dotted lines correspond to the closed system ($\gamma^0_i=0$). Lightblue/yellow lines correspond to real/imaginary parts of $\epsilon^\pm_A$, whereas darkblue/red ones do to $\epsilon^\pm_B$. Dashed/solid lines distinguish $+/-$ states. \textbf{Right:} Supercurrent as a function of $\phi$ for different temperatures. Black curve corresponds to the Hermitian analog at $T=10^{-8}$. Light/dark green dotted lines mark the position of the EPs in both panels. System parameters are fixed as $\Delta-t=0.2$, $t_{23}=\Delta=1$, $\gamma^0_2=1$, $\gamma^0_3=0.3$ and $\gamma^0_1=\gamma^0_4=0$.}
    \label{fig:4Majorana-twochains_spectrum}
\end{figure}
\begin{figure}[ht]
    \centering
    \includegraphics[width=0.45\linewidth]{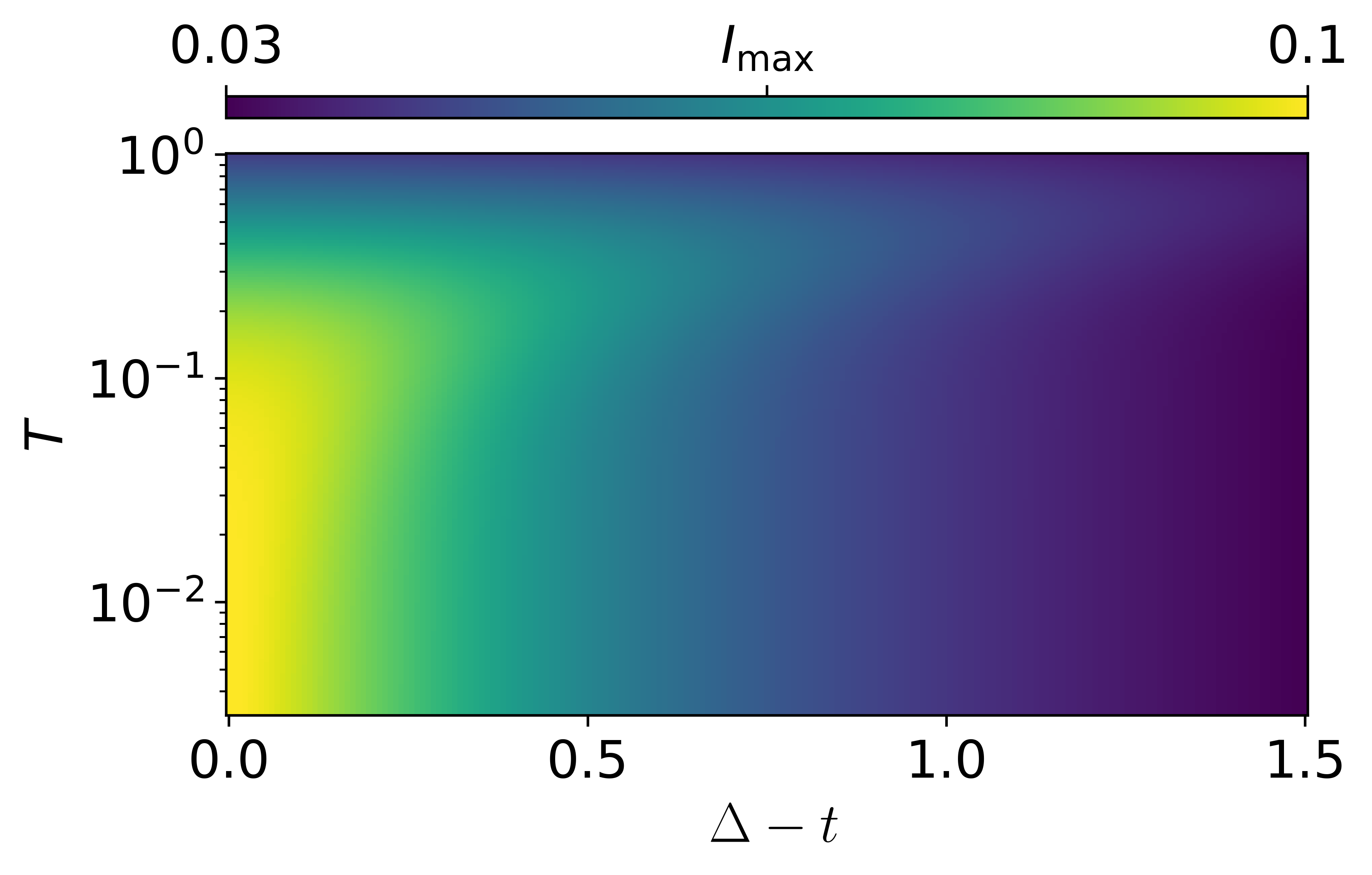}
    \caption{\textbf{Maximum supercurrent $I_\mathrm{max}$} as a function of $T$ and $\Delta-t$. System parameters are fixed for the same values as in Fig. \ref{fig:4Majorana-twochains_spectrum}.}
    \label{fig:I-max}
\end{figure}

\subsection{On the zero-temperature limit}

We can conclude that for the configurations that we have studied for both ABS and four-Majorana models, the supercurrent converges to a finite, cutoff-independent value (\ref{eq:SI_T-limit_supercurrent_ABS}) when $T\to 0$, although from (\ref{eq:T-limit_supercurrent}) this could not be true in general. However, we can demonstrate that the cutoff-dependence in (\ref{eq:T-limit_supercurrent}) cancels exactly for every case. Indeed, given a generic Hamiltonian $H=H_0 - i\mathbf{\gamma}_0$, its trace is a conserved quantity under unitary transformations. Thus, we can write
\begin{equation}
    \sum_j \Im(\epsilon_j) = -\Tr(\mathbf{\gamma}_0) \quad\Rightarrow\quad \sum_j \frac{\partial\Im(\epsilon_j)}{\partial\phi} = -\Tr(\partial_\phi\mathbf{\gamma}_0)
\end{equation}
Hence, supposing a phase-independent $\mathbf{\gamma}_0$ matrix, then the previous sum will be constant with respect $\phi$ and the term
\begin{equation}
    -\frac{T}{D}\log\Gamma\left(\frac{1}{2}+\frac{D}{2\pi T}\right) \sum_j\frac{\partial\gamma_j}{\partial\phi} = 0
\end{equation}
for any temperature, leading to the general $T\to 0$ limit
\begin{equation} \label{eq:T-limit_supercurrent_general}
    I_{T\to0} = \frac{1}{2}\sum_j\left[\frac{1}{\pi}\arctan\left(\frac{\gamma_j}{E_j}\right)\frac{\partial E_j}{\partial\phi} + \frac{1}{\pi}\left(\log\sqrt{\gamma_j^2+E_j^2} - 1\right)\frac{\partial\gamma_j}{\partial\phi}\right]
\end{equation}
Note that this justification could be not true for a phase-dependent $\mathbf{\gamma}_0$ matrix.

\section{Entropy}\label{sec:Appendix_D}
Defining the entropy as $S=-\partial F/\partial T$, from (\ref{eq:SI_free-energy}), in the $D\to\infty$ limit we have
\begin{equation} \label{eq:entropy}
\begin{aligned}
   S =& \log(2\pi) - 2\Re\log\Gamma\left(\frac{1}{2}+\frac{i\gamma-E}{2i\pi T}\right) + \frac{\gamma}{\pi T}\Re\psi\left(\frac{1}{2}+\frac{i\gamma-E}{2i\pi T}\right) - \frac{E}{\pi T}\Im\psi\left(\frac{1}{2}+\frac{i\gamma-E}{2i\pi T}\right)
   \\
   & + \frac{2\gamma}{D}\log\Gamma\left(\frac{1}{2}+\frac{D}{2\pi T}\right) - \frac{\gamma}{\pi T}\psi\left(\frac{1}{2}+\frac{D}{2\pi T}\right)
\end{aligned}
\end{equation}
Alternatively, from the approximated expression (\ref{eq:free-energy_approx}),
\begin{equation} \label{eq:entropy_approx}
    S \approx \log(2\pi) - 2\Re\log\Gamma\left(\frac{1}{2}+\frac{i\gamma-E}{2i\pi T}\right) + \frac{\gamma}{\pi T}\Re\psi\left(\frac{1}{2}+\frac{i\gamma-E}{2i\pi T}\right) - \frac{E}{\pi T}\Im\psi\left(\frac{1}{2}+\frac{i\gamma-E}{2i\pi T}\right)
\end{equation}

Fig. \ref{fig:entropy} shows entropy as a function of $T$ for both expressions (\ref{eq:entropy}) and (\ref{eq:entropy_approx}). This last approximated form leads to divergences in the entropy at low $T$, so the cutoff becomes necessary. This can be understood by the analytical asymptotics of both expressions,
\begin{equation}
\begin{aligned}
    S_{T\to 0} &= \log\frac{\sqrt{\gamma^2+E^2}}{2\pi T} - \frac{\gamma}{D}\log\frac{D}{2\pi T} - 1
    \\
    S_{T\to 0} &\approx \log\frac{\sqrt{\gamma^2+E^2}}{2\pi T} + \frac{\gamma}{\pi T} - 1
\end{aligned}
\end{equation}
The full limit goes to a finite value for large $D$, whereas the approximated form shows a divergence $\sim\log 1/T$.

\begin{figure}[ht]
    \centering
    \includegraphics[width=0.8\linewidth]{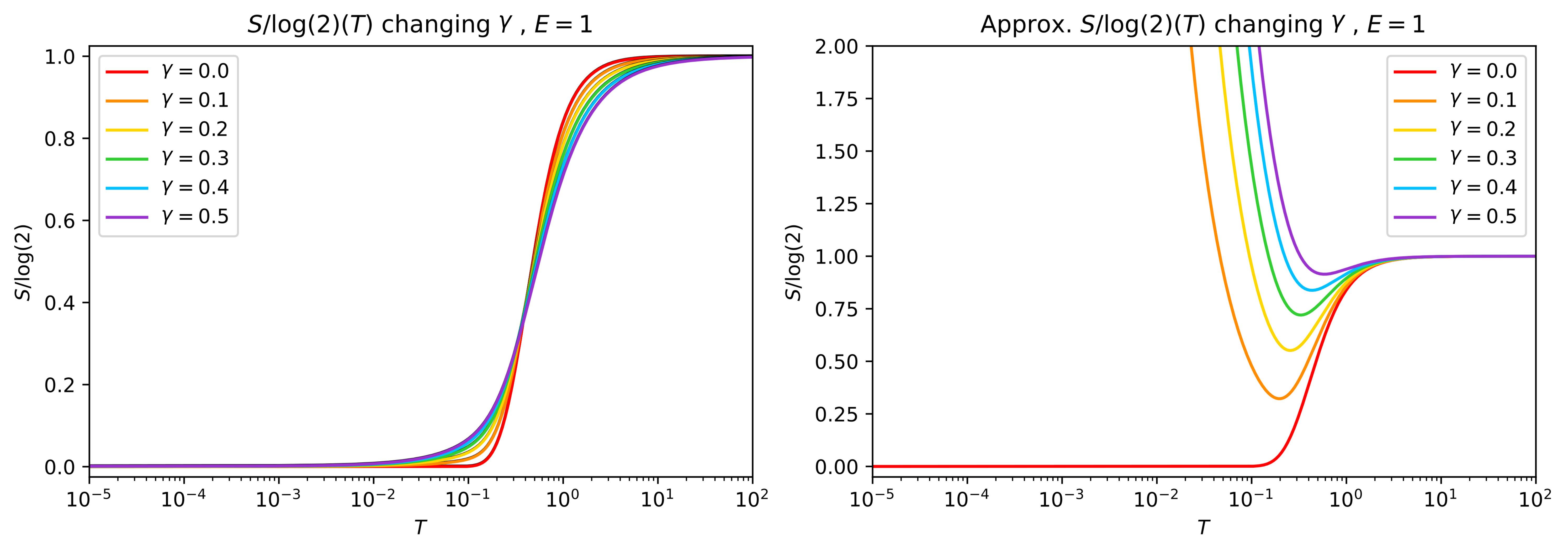}
    \caption{\textbf{Evolution of entropy} as a function of $T$. \textbf{Left:} full expression (\ref{eq:entropy}). \textbf{Right:} approximated expression (\ref{eq:entropy_approx}).}
    \label{fig:entropy}
\end{figure}

\subsection{Entropy in BdG systems: role of EPs}
Our formalism allows us to readily explain fractional plateaus of entropy in Eq. (\ref{eq:entropy}) as different contributions $S^+\neq S^-$, within temperature windows where only an odd number of states contribute to the total entropy. These plateaus, reported in e.g. Refs. \cite{SelaPRL:19,Smirnov} can be understood in a very elegant manner as originated from EPs in the complex spectrum. We illustrate this idea in Fig. \ref{fig:entropy-EP}, where we plot the entropy steps before and after an EP, showing that the appearance of a fractional plateau is due to the different contributions $S^+$ and $S^-$ of the states $\epsilon^+$ and $\epsilon^-$.

\begin{figure}[ht]
    \centering
    \includegraphics[width=0.7\linewidth]{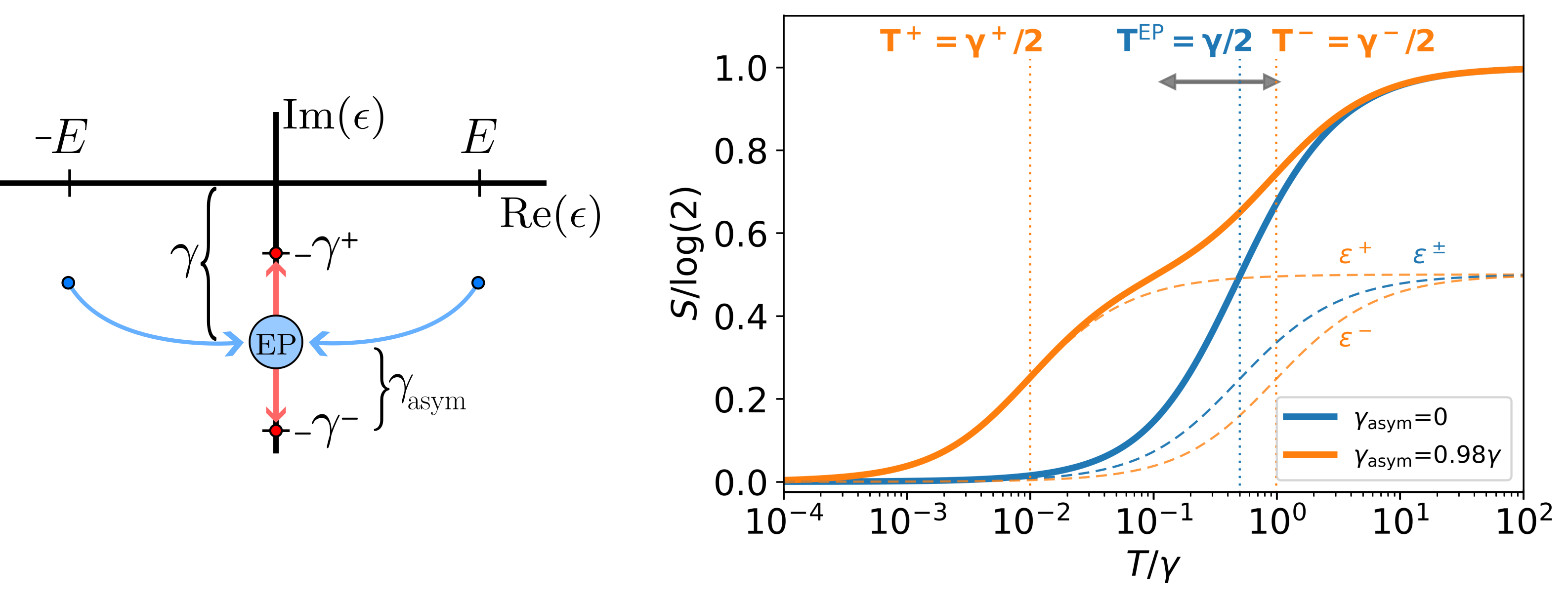}
    \caption{\textbf{Entropy development after an exceptional point}. Development of a fractional entropy plateau after the coalescence of two BdG poles. \textbf{Blue:} before the EP, real and imaginary parts of both eigenvalues have identical magnitude, $\epsilon^{\pm}=\pm E -i\gamma$, and thus both poles contribute to the entropy (\ref{eq:entropy}) at the same temperature until the EP is reached ($T^{\mathrm{EP}}=|\epsilon^\pm|/2=\gamma/2$, blue dotted curve). \textbf{Orange:} after the EP, $E=0$ and their imaginary parts are no longer identical, $\epsilon^\pm=-i\gamma^\pm$, with $\gamma^- - \gamma^+=2\gamma_\mathrm{asym}$. Hence, their contributions to the entropy come at different temperatures ($T^{\pm}=\gamma^\pm/2$, orange dotted curves), giving rise to a fractional plateau $S=\log(2)/2$ of width $\gamma_\mathrm{asym}$. Dashed curves refer to each single-state contribution, whereas solid lines correspond to the total entropy of the system.}
    \label{fig:entropy-EP}
\end{figure}

\section{Generalized Maxwell relation}\label{sec:Appendix_E}

Regarding the detection of these fractional entropy plateaus, we start with the definition of the Helmholtz free energy
\begin{equation}
    F = -S \, dT - N \, d\mu + I \, d\phi + \dots
\end{equation}

One can extract from this expression a differential Maxwell relation between the entropy and another thermodynamic function $y$ by continuously changing their conjugate variables $T$ and $x$, respectively, such as
\begin{equation}
    \frac{\partial S}{\partial x} = - \frac{\partial y}{\partial T}
\end{equation}
or, written in an integral form,
\begin{equation}
    \Delta S_{x_1\to x_2} = -\int_{x_1}^{x_2} \frac{\partial y}{\partial T} \, dx
\end{equation}

There has been well-established progress in entropy measurement via Maxwell relations involving occupation, $y(x)=-N(\mu)$. We propose a novel application of this procedure, employing instead the Josephson supercurrent, $y(x)=I(\phi)$. Hence, integrating $\partial I/\partial T$ between $\phi_1=0$ and $\phi_2=\pi$, allows us to get entropy changes of a mesoscopic system from current measurements and thus validate the very precise predictions about fractional entropy plateaus presented in this work.

Note that we have used this same procedure to compute the free energy, the current, and the entropy. Indeed, starting from the integral form of the occupation, which has a simple pole structure, allows us to get these macroscopic quantities by just taking derivatives. Here we would like to note that all the above derivations assume that $\rho(\omega)$ does not depend on $T$ or $\mu$, e.g. self-consistently through the occupation itself, something that is clearly not true in interacting systems (e.g. an Anderson model with Kondo effect). A derivation along the previous lines considering interactions explicitly deserves further investigation.

\section{Temperature effects}\label{sec:Appendix_F}

While in the minimal Kitaev chain model presented in this work the pairing potential $\Delta$ is supposed to be a temperature-independent parameter, in a real experiment it comes from crossed Andreev reflections mediated by a middle segment separating both quantum dots. Such segment is typically a semiconducting region proximitized by a superconductor such that $\Delta$ is much lower than the true superconducting gap (in the experiments of Ref. ~\cite{Dvir-Nature2023,haaf2024engineering} $\Delta$ is around $10\mu$eV while the gap of the parent aluminum superconductor is around $270\mu$eV. Thus, a regime where $T\gg\Delta$ without destroying superconductivity is possible.  
Of course, there is no need to reach such a large-temperature regime: Majorana physics (evidence of fractional entropy plateaus) always occurs at lower temperatures. In this section, we include the temperature dependence of the parent superconducting gap in a DQD model to understand this argument fully.

\subsection{Temperature dependence of the parent superconducting gap \label{Gap-T}}
The self-consistent integral formula of the superconducting gap at finite temperature takes the form \cite{Bruus-Flensberg}
\begin{equation}
\label{gap-T}
    \eta = \int_0^{\omega_D} d\xi \, \frac{\tanh\left(\frac{1}{2T}\sqrt{\xi^2+\Delta_T^2}\right)}{\sqrt{\xi^2+\Delta_T^2}}
\end{equation}
where $\eta=\frac{1}{N(0)V}$ is the (inverse) interaction strength, $\omega_D$ is the Debye frequency of the superconductor and $\Delta_T$ its superconducting gap at a finite temperature $T$. In the case of Aluminum, $\eta=1/0.18$ and $\omega_D=375k_B\approx 0.032$ eV (in natural units) \cite{de-Gennes}. With these quantities, the limit $\Delta_T\to 0$ allows us to extract the critical temperature of the superconducting material:
\begin{equation}
\label{critical-aluminium}
    T_c = \frac{2e^\gamma}{\pi} \omega_D e^{-\eta} \approx 141.651  \; \mu\mathrm{eV}
\end{equation}
where $\gamma$ is the Euler's constant. On the other hand, with the limit $T\to 0$ we can calculate the gap at zero temperature:
\begin{equation}
    \Delta_0 = \lim_{T\to 0}\Delta_T = \omega_D \sinh^{-1}\left[ \log\left( \frac{2e^\gamma}{\pi}\frac{\omega_D}{T_c}\right) \right] \approx 249.858 \; \mu\mathrm{eV}
\end{equation}

Then, with all of these quantities, we can (numerically) calculate the temperature dependence of the superconducting gap, as depicted in Fig. \ref{fig:gap-T}.

\begin{figure}[ht]
    \centering
    \includegraphics[width=0.45\linewidth]{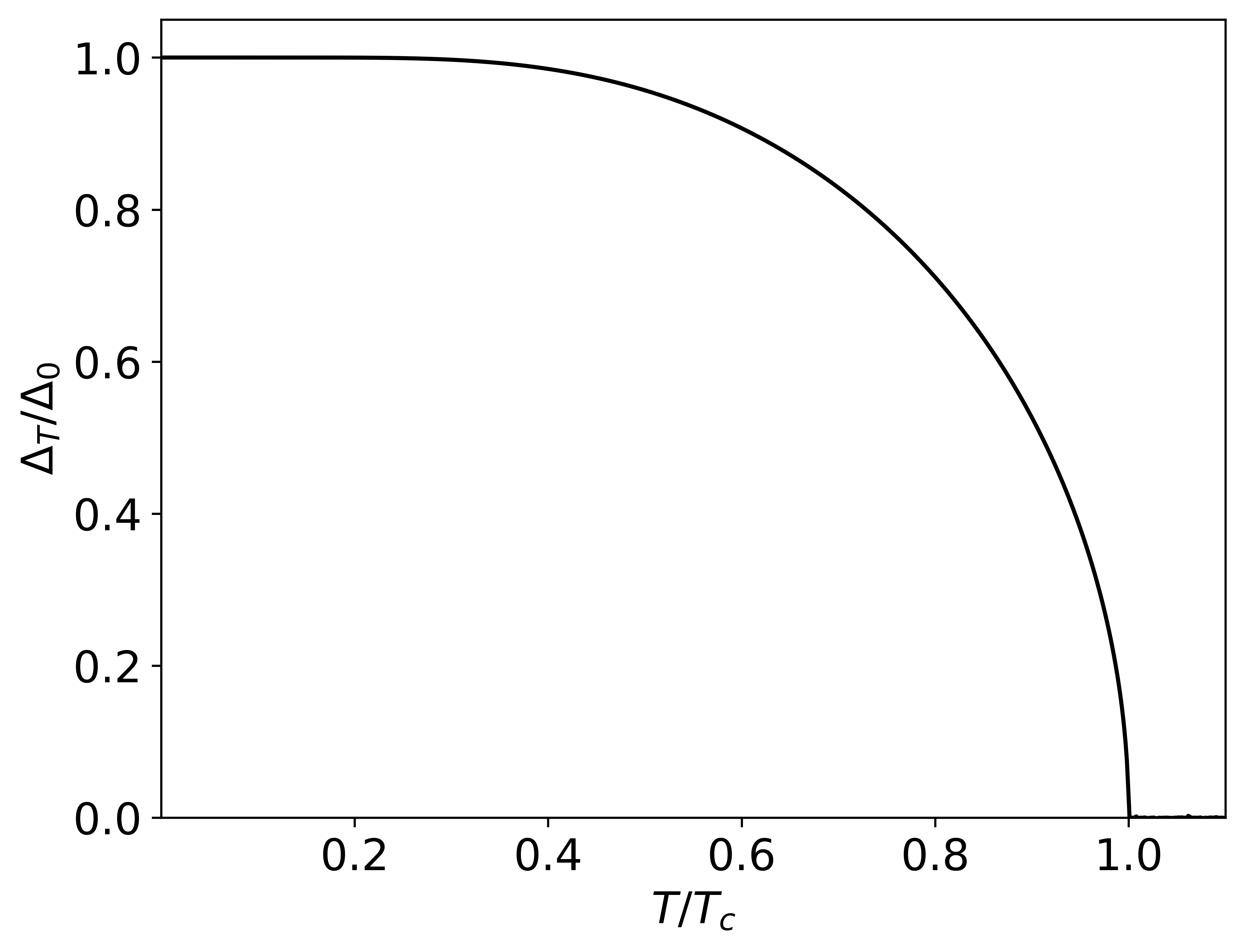}
    \caption{\textbf{Temperature dependence of the superconducting gap}.}
    \label{fig:gap-T}
\end{figure}
\subsection{Microscopic model of a minimal Kitaev chain \label{microscopic-hybrid}}
Now, we discuss a realistic realization of the minimal Kitaev model based on quantum dots. As discussed in the main text, the simplest setup consists of a double quantum dot (DQD) model containing an interdot SC pairing interaction $\Delta$ and a single particle hopping term $t$. This DQD model is described by the Hamiltonian
\begin{equation}
\label{minimal-chain}
    H_\mathrm{DQD} = -\mu_L c_L^\dagger c_L - \mu_R c_R^\dagger c_R - t c_L^\dagger c_R + \Delta c_L c_R + \mathrm{H.c.}
\end{equation}
where $c^\dagger_{L/R}$ ($c_{L/R}$) denote creation (annihilation) operators on the left/right QD with a chemical potential $\mu_{L/R}$.  If, additionally, both QDs are coupled to normal reservoirs with rates $\gamma_1^0$ and $\gamma_2^0$, this system is a realization of a Non-Hermitian minimal Kitaev chain containing Majorana modes, whose eigenenergies at $\mu_1=\mu_2=0$ can be expressed analytically,
\begin{equation}
\label{eigenvalues}
\begin{aligned}
    \epsilon_\mathrm{outer}^\pm = -i\frac{\gamma_1^0+\gamma_2^0}{2} \pm \sqrt{(t-\Delta)^2 - (\gamma_1^0-\gamma_2^0)^2/4}
    \\
    \epsilon_\mathrm{inner}^\pm = -i\frac{\gamma_1^0+\gamma_2^0}{2} \pm \sqrt{(t+\Delta)^2 - (\gamma_1^0-\gamma_2^0)^2/4}
\end{aligned}
\end{equation}
showing an exceptional point at $|t-\Delta|=|\gamma_1^0-\gamma_2^0|/2$ for the outer states and at $|t+\Delta|=|\gamma_1^0-\gamma_2^0|/2$ for the inner ones.

Interestingly, the above model can be realized microscopically by means of a common superconductor-semiconductor hybrid region which couples both quantum dots. Specifically, the QDs couple via a subgap Andreev bound state (ABS) living in the middle region. Such ABS coupling allows for crossed Andreev reflection (CAR) and single-electron elastic co–tunneling (ECT), with coupling strengths $\Delta$ and $t$, respectively, which can be identified with the parameters in Eq. \eqref{minimal-chain}. Physically, the spin-orbit coupling in semiconductor-superconductor provides a spin-mixing term for tunneling electrons. The spin-mixing term can lead to finite ECT and CAR amplitudes for spin-polarized QDs when the spin-orbit and magnetic fields are non-colinear.  This ABS provides a low-excitation energy for ECT and CAR, therefore dominating the coupling between the two QDs. Starting from a microscopic model of two quantum dots connected by a semiconductor-superconductor
middle region, one can apply a Schrieffer-Wolff transformation to obtain effective cotunneling-like couplings of the form \cite{PhysRevLett.129.267701}
\begin{equation}
\label{eq:SI_CAR-ECT}
\begin{aligned}
    t & \sim \frac{t_L t_R}{\Delta_0} \left(\frac{2uv}{E_\mathrm{ABS}/\Delta_0}\right)^2
    \\
    \Delta & \sim \frac{t_L t_R}{\Delta_0} \left(\frac{u^2-v^2}{E_\mathrm{ABS}/\Delta_0}\right)^2
\end{aligned}
\end{equation}
where $t_{L/R}$ are the local tunneling strengths between each QD and the central region and 
\begin{equation}
    E_\mathrm{ABS}=\Delta_0\sqrt{z^2 + 1}
\end{equation}
is the energy of the Andreev state, with $z\equiv\mu_\mathrm{ABS}/\Delta_0$ being the chemical potential of the middle region, and BdG coefficients given by $u^2=1-v^2=1/2 + \mu_\mathrm{ABS}/(2E_\mathrm{ABS})$. 

As discussed in Ref.~\cite{PhysRevLett.129.267701}, the crossed Andreev reflection  $\Delta$ has a single peak centered at centered at $z = 0$ and decays as $z^{-4}$ at large $|z|$, while elastic cotunneling $t$ has double peaks located at $z = \pm 1$ and decays as $z^{-2}$ at large $|z|$. It also has an exact dip at $z = 0$ due to destructive interference. The precise energy dependence of $\Delta$ and $t$ allows us to vary their relative amplitudes by changing the chemical potential $\mu_\mathrm{ABS}$ of the ABS and tune them to precise points, so-called "sweet spots", where $\Delta=t$, for an example see Fig. \ref{fig:Delta-t-mu}.
\begin{figure}[ht]
    \centering
    \includegraphics[width=0.45\linewidth]{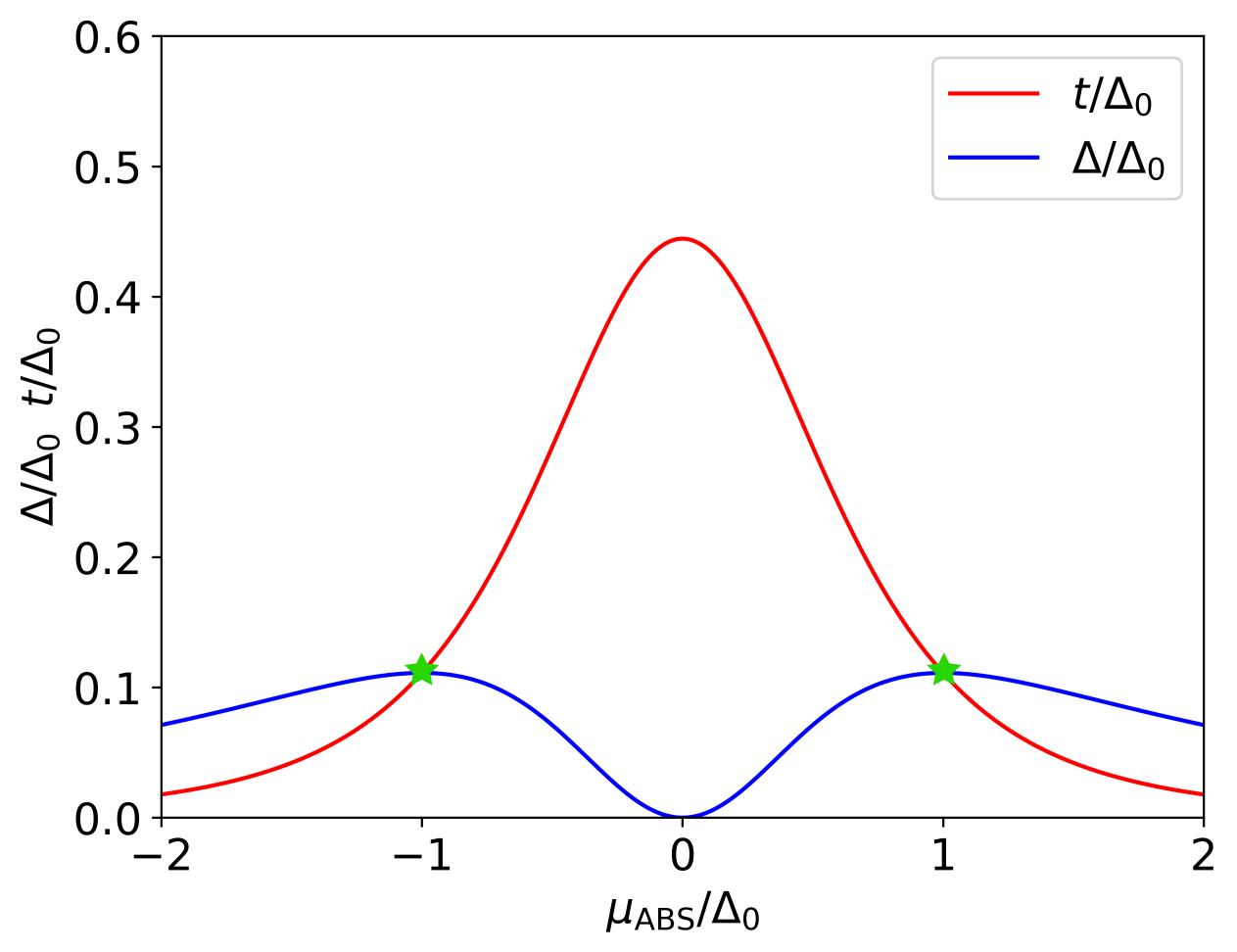}
    \caption{\textbf{Couplings strengths $\Delta$ and $t$} as a function of the chemical potential of the central region $\mu_\mathrm{ABS}$ at $T=0$. Green stars mark the position of the sweet spot $t=\Delta$ at $\mu_\mathrm{ABS}=\pm\Delta_0$. Parameters used: $t_L=t_R=2/3\Delta_0$.}
    \label{fig:Delta-t-mu}
\end{figure}
Here, we would like to emphasize that such ABS-mediated coupling of CAR and ECT terms using a middle semiconducting region with Rashba interaction can be modified experimentally by gate voltages. Indeed, precise tuning of the relative amplitudes between the two processes  \emph{has already been demonstrated in different experimental platforms and configurations}
\cite{Dvir-Nature2023,haaf2024engineering,haaf2024edgebulkstatesthreesite}. Relevant to our proposal is the fact that the latter experiment has demonstrated the precise control of a three-site Kitaev chain by applying a superconducting phase difference,  an experimental breakthrough that paves the way towards Josephson junctions based on minimal Kitaev chains of quantum dots like the ones we discuss in our paper.

\subsection{Temperature dependence of CAR and ECT: impact on Exceptional Points}
\begin{figure}[ht]
    \centering
    \includegraphics[width=0.9\linewidth]{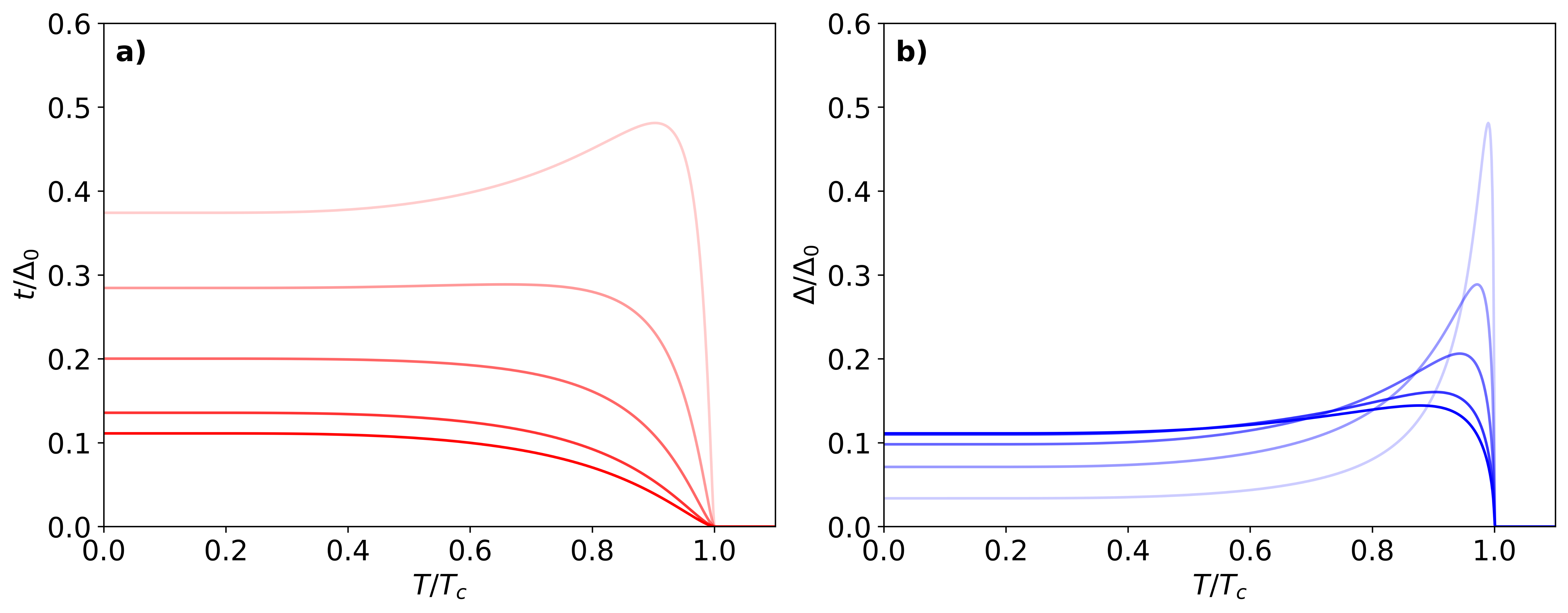}
    \caption{\textbf{Couplings strengths $\Delta$ and $t$}. Temperature dependence of $t$ and $\Delta$ for different values of $\mu_\mathrm{ABS}/\Delta_0=0.3, 0.5, 0.7, 0.9, 1$ (increasing opacity of the curves). Parameters used: $t_L=t_R=2/3\Delta_0$.}
    \label{fig:Delta-t-T}
\end{figure}
In the above derivation, which closely follows that of Ref. \cite{PhysRevLett.129.267701}, the gap $\Delta_0$ of the parent superconductor, was assumed to be temperature-independent. We now extend this microscopic theory and explicitly consider the temperature dependence of the Aluminium parent gap according to Eq. \eqref{gap-T}. This makes the crossed Andreev
reflection and elastic cotunneling terms in Eq. \eqref{eq:SI_CAR-ECT} temperature-dependent too by just making the substitution $\Delta_0\rightarrow \Delta_T$. We show their resulting temperature dependence in Figs. \ref{fig:Delta-t-T}a and b. Interestingly, their temperature dependence is highly nonmonotonic, with large regions where they remain nearly constant until they decrease near $T_c$ where the parent gap closes. This region near $T_c$ also develops a softened singularity, reminiscent of a BCS-like peak, as the chemical potential approaches $\mu_\mathrm{ABS}/\Delta_0\approx 1$.

\begin{figure}[ht]
    \centering
    \includegraphics[width=0.9\linewidth]{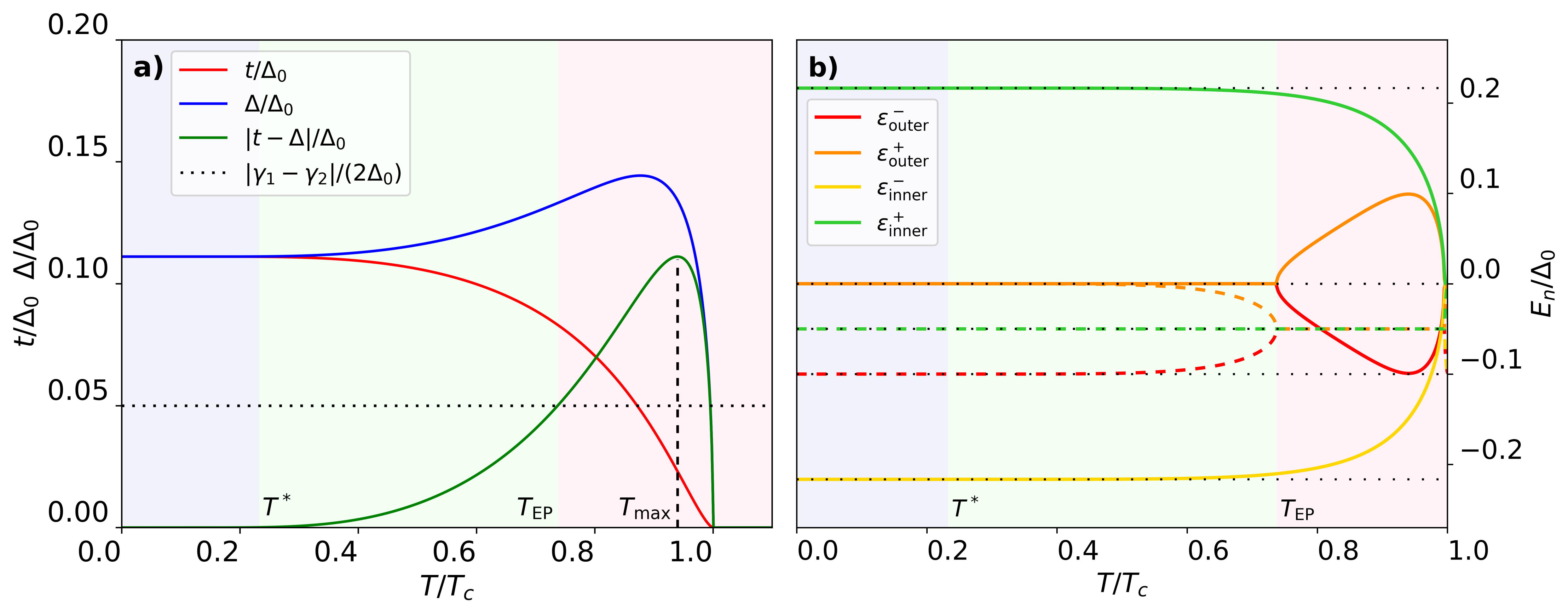}
    \caption{\textbf{(a) Temperature dependence of $t$ and $\Delta$}. Lilac/green regions border corresponds to $T^*$, such that $|t-\Delta|=10^{-4}\Delta_0$. Green/pink regions border corresponds to $T_\mathrm{EP}$, where appears the EP bifurcation between outer states ($|t-\Delta|=|\gamma_1^0-\gamma_2^0|/2$). The maximum value of $|t-\Delta|$ is marked with a vertical dashed line at $T_\mathrm{max}$. \textbf{(b) Complex energy spectrum of the system}. Solid/dashed lines correspond to the real/imaginary parts of $\epsilon_\mathrm{inner/outer}^\pm$. 
    Black dotted lines correspond to the $T\to 0$ approximation of Eq. (\ref{eq:SI_T-0_approx}), which, for this choice of parameters, agree with $T$-finite results until $T^*\approx 0.23T_c$ (lilac region in both figures defined as $|t-\Delta|\leq 10^{-4}\Delta_0$).  Parameters used: $\gamma_1^0=0$ and $\gamma_2^0=0.1\Delta_0$; $\mu_\mathrm{ABS}/\Delta_0=1$; $t_L=t_R=2/3\Delta_0$.}
    \label{fig:spectrum-T}
\end{figure}

We now include the explicit temperature dependencies of $\Delta$ and $t$ and show de complex spectrum given by Eq. \eqref{eigenvalues}, see Fig. \ref{fig:spectrum-T}. We first fix $\mu_\mathrm{ABS}=\Delta_0$, where the system reaches the sweet spot at zero temperature, namely $t=\Delta$ by using $\Delta_0$ in Eq. \eqref{eq:SI_CAR-ECT}, and study the temperature dependence of the difference $|t-\Delta|$ as $\Delta_0\rightarrow \Delta_T$ (Fig. \ref{fig:spectrum-T}a). The temperature evolution of $|t-\Delta|$ governs the emergence of exceptional point (EP) bifurcations, which are now temperature-dependent (Fig. \ref{fig:spectrum-T}b). As the temperature increases in Fig. \ref{fig:spectrum-T}a, both couplings remain equal $|t-\Delta|=0$ until reaching a temperature $T^*$ (lilac regions). Above $T^*$ 
both couplings start to differ $|t-\Delta|\neq 0$.
Nevertheless, as long as the condition $|t-\Delta|<|\gamma_1-\gamma_2|/2$ is fulfilled, the system still hosts a pair of Majorana zero modes (green region). This can be explicitly seen in Fig. \ref{fig:spectrum-T}b where we plot the full temperature-dependent evolution of the complex spectrum. Specifically, the green region corresponds to a regime where two Majorana zero modes ($\Re(\epsilon_\mathrm{outer}^+)=\Re(\epsilon_\mathrm{outer}^-)=0$) are asymmetrically coupled to the reservoir ($\Im(\epsilon_\mathrm{outer}^+)\neq \Im(\epsilon_\mathrm{outer}^-)\neq 0$). This finite-temperature regime 
\emph{with Majorana zero modes} persists until $|t-\Delta|$ is large enough to close the bifurcation. Specifically, the exact value of the temperature at which the EP forms, $T_\mathrm{EP}$, is determined by the point where the condition $|t-\Delta|=|\gamma_1^0-\gamma_2^0|/2$ is fulfilled
\begin{equation}
    |t-\Delta| = \frac{t_Lt_R\Delta_T|\Delta_T^2-\mu_\mathrm{ABS}^2|}{(\Delta_T^2+\mu_\mathrm{ABS}^2)^2}= \frac{|\gamma_1^0-\gamma_2^0|}{2} \;,
\end{equation}
which depends on the explicit form of the temperature-dependent gap $\Delta_T$.
Above $T_\mathrm{EP}$, namely $|t-\Delta|>|\gamma_1^0-\gamma_2^0|/2$ the system no longer contains Majorana zero modes (pink region). The lilac region, on the other hand, should be understood as the temperature $T^*$ below which the $T\to 0$ limit is valid (in terms of EPs, this corresponds to $\Im [\epsilon_\mathrm{outer}^+]\to 0$). In this $T\to 0$ ($\Delta_T\to\Delta_0$) limit, the poles can be written as
\begin{equation}\label{eq:SI_T-0_approx}
\begin{aligned}
    \epsilon_\mathrm{outer}^\pm = -i\frac{\gamma_1^0+\gamma_2^0}{2} \pm i \frac{|\gamma_1^0-\gamma_2^0|}{2} = -i\gamma_{1/2}^0
    \quad,\quad \epsilon_\mathrm{inner}^\pm = -i\frac{\gamma_1^0+\gamma_2^0}{2} \pm \sqrt{\frac{t_L^2t_R^2}{4\Delta_0^2} - \frac{(\gamma_1^0-\gamma_2^0)^2}{4}} \;.
\end{aligned}
\end{equation}
where we choose $\max(\gamma_1^0,\gamma_2^0)=|\epsilon_\mathrm{outer}^-|>|\epsilon_\mathrm{outer}^+|=\min(\gamma_1^0,\gamma_2^0)$ for convenience, such that $\epsilon_\mathrm{outer}^{-(+)}$ always refers to the pole most (least) coupled to the reservoir.

For the choice of parameters in Fig. \ref{fig:spectrum-T}b, we find an estimate of $T^*\approx 0.23T_c$ (although the precise value of $T^*$ is somewhat arbitrary when performing numerics, we here use a conservative lower bound of $|t-\Delta|\leq 10^{-4}\Delta_0$ to draw the lilac boundary, for comparison we also plot the analytical values in Eq. \eqref{eq:SI_T-0_approx}, see black dotted lines in Fig. \ref{fig:spectrum-T}b). 

Even though the calculations in Fig. \ref{fig:spectrum-T} have been performed for the particular (but reasonable) values of $t_{L/R}=2/3\Delta_0$ and $|\gamma_1^0-\gamma_2^0|=0.1\Delta_0$, we can extract some general conclusions that enable the possibility of an experimental demonstration of our claim: although the magnitude of $|t-\Delta|$ depends on the local tunneling strengths $t_{L/R}$, their functional form against the temperature remains the same, reaching a maximum value of
\begin{equation}
    |t-\Delta|_\mathrm{max} = \frac{t_Lt_R}{4\mu_\mathrm{ABS}}
\end{equation}
at $\Delta_T=(\sqrt{2}-1)\mu_\mathrm{ABS}$, which is independent of $\gamma_{1/2}^0$ and $t_{L/R}$, and corresponds to a large value of $T_\mathrm{max}\approx 0.94T_c$ when $\mu_\mathrm{ABS}=\Delta_0$. Conversely, $T_\mathrm{EP}$ depends on the ratio between the reservoir coupling asymmetry $|\gamma_1^0-\gamma_2^0|$ and the tunneling strengths $t_{L/R}$, and it must be smaller than $T_\mathrm{max}$ such that the EP can develop. For the choice of parameters in Fig. \ref{fig:spectrum-T}, $T_\mathrm{EP}\approx 0.74T_c<T_\mathrm{max}$. Particularly, for values of $|\gamma_1^0-\gamma_2^0|/2$ greater than $|t-\Delta|_\mathrm{max}=\frac{t_Lt_R}{4\Delta_0}$ ($\mu_\mathrm{ABS}=\Delta_0$), the system would never leave the non-trivial topological phase. By rewriting the couplings in terms of the parent gap as $t_L=\eta_L\Delta_0$ and $t_R=\eta_R\Delta_0$, this condition reads $|\gamma_1^0-\gamma_2^0|>\eta_L\eta_R\Delta_0/2$. Since in realistic settings both $\eta_L,\eta_R\ll 1$, this implies that there is no need of a huge coupling asymmetry for keeping the existence of Majorana zero modes even for temperatures close to $T_c$.

\subsection{Temperature dependence of CAR and ECT: impact on entropy calculations}

Armed with the above results, we finally calculate the entropy of a minimal Kitaev chain including the temperature dependence of the effective parameters $\Delta$ and $t$, and hence of the eigenvalues that bifurcate. 
By tuning $\mu_\mathrm{ABS}=\Delta_0$, the system approaches the sweet-spot regime for temperatures below $T_\mathrm{EP}$, where the poles $\epsilon_\mathrm{outer}^\pm$ exhibit an EP, as shown in Fig. \ref{fig:spectrum-T}b. Indeed, as we explained in the previous section, for $T<T_\mathrm{EP}$ these outer poles correspond to two Majorana zero modes that are asymmetrically coupled to the reservoir.

On the other hand, as explained in the main text of this work, the entropy jumps associated with these poles occur at the crossover temperatures $T_\mathrm{outer}^\pm = |\epsilon_\mathrm{outer}^\pm|/2$. If the magnitudes of the poles differ, $|\epsilon_\mathrm{outer}^-|\neq|\epsilon_\mathrm{outer}^+|$, a fractional plateau $S = \log(2)/2$ appears, and its width is given by $T_\mathrm{outer}^- - T_\mathrm{outer}^+$. Thus, this fractional plateau is observable for temperatures below $T_\mathrm{outer}^-$. Importantly, this temperature is not directly given by $T_\mathrm{EP}$: although the system hosts MZMs up to $T_\mathrm{EP}$, the fractional plateau only emerges when the mode most coupled to the reservoir has not yet contributed to the entropy, which happens at $T_\mathrm{outer}^-$. Here, it is important to emphasize that since $\epsilon_\mathrm{outer}^-(T)$ depends on the temperature (Fig. \ref{fig:spectrum-T}b), the value of this crossover temperature becomes a self-consistent problem,
\begin{equation}
    T_\mathrm{outer}^- = \frac{|\epsilon_\mathrm{outer}^-(T_\mathrm{outer}^-)|}{2}.
\end{equation}
Thus, $T_\mathrm{outer}^-$ will be given by the crossing point between the curve $|\epsilon_\mathrm{outer}^-(T)|/2$ and the line $y(T)=T$ (and the same for the rest of poles), see Fig. \ref{fig:new-entropy}a. Importantly, in this figure all the crossover temperatures lie in the lilac region, where the limit $T\to 0$ is valid, Eq. \eqref{eq:SI_T-0_approx}. Under this limit, we have in general $T_\mathrm{outer}^-=\max(\gamma_1^0,\gamma_2^0)/2$, being completely valid as long as $\gamma_1^0,\gamma_2^0<2T^*$.

Specifically, for the particular choice of parameters in Fig. \ref{fig:spectrum-T} (cyan line of Fig. \ref{fig:new-entropy}b for $\mu_\mathrm{ABS}=\Delta_0$), this temperature is one order of magnitude smaller than $T_c$ ($T_\mathrm{outer}^-=0.088T_c<T^*$), so that the $T\to 0$ approximation is valid for this and the rest of the poles, Fig. \ref{fig:new-entropy}a. For the parameters used in Eq. \eqref{critical-aluminium} it would give a temperature of the order of $T_\mathrm{outer}^-\approx 15\mu\mathrm{eV}$ (which in real temperature units corresponds to a temperature of $T_\mathrm{outer}^-\approx 170mK$ well within the experimental range of observability) where the fractional plateau (cyan line for $\mu_\mathrm{ABS}=\Delta_0$) develops. Importantly, typical gaps observed in the experiments demonstrating minimal Kitaev chains based on quantum dots ~\cite{Dvir-Nature2023,haaf2024engineering,haaf2024edgebulkstatesthreesite} are $\Delta\approx 10-20\mu$eV, in the same energy range of our estimation of the temperature range below which the fractional plateau in the entropy emerges. This supports our claim that unreasonably low temperatures $T\ll \Delta$ are \emph{not needed} to observe our predicted fractional plateaus.

\begin{figure}[ht]
    \centering
    \includegraphics[width=0.9\linewidth]{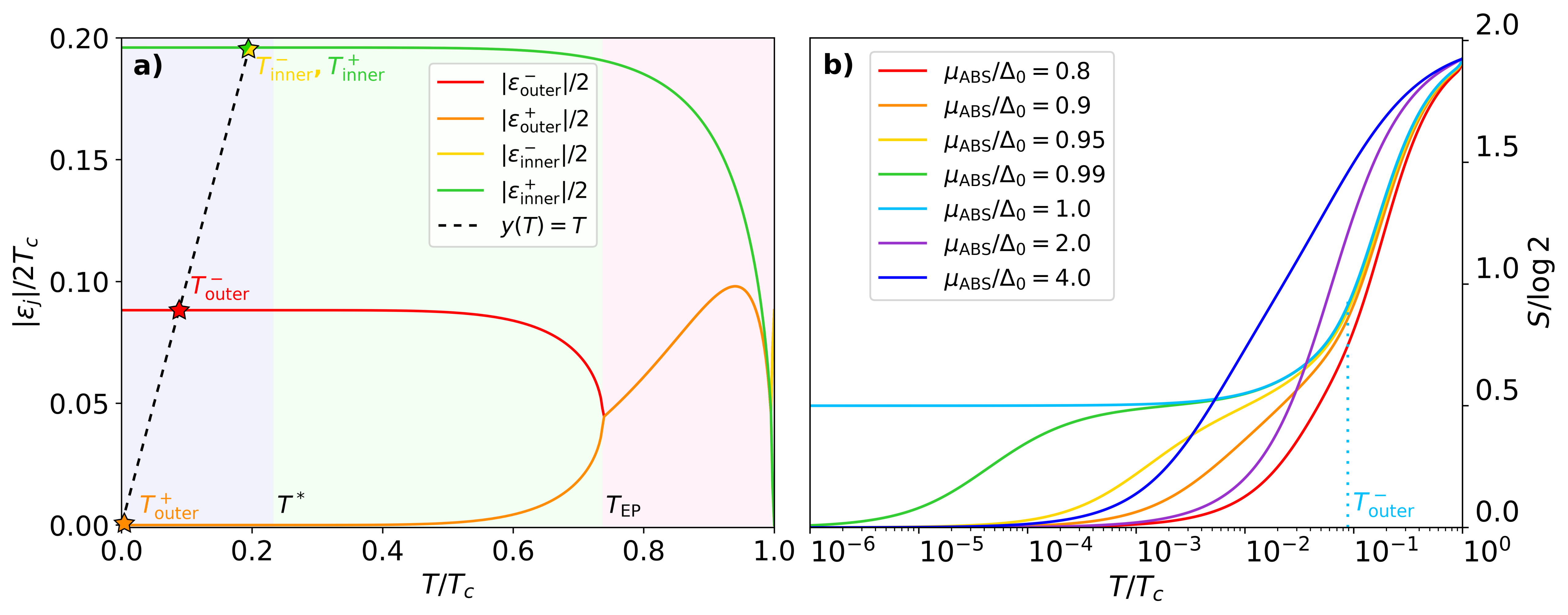}
    \caption{\textbf{(a) Crossover temperatures}. Temperature dependence of the absolute value of the complex poles for $\mu_\mathrm{ABS}=\Delta_0$. Colored stars mark the points where $T=|\epsilon_j(T)|/2$, corresponding to the crossover temperatures of each pole. \textbf{(b) Entropy}. Temperature dependence of entropy for different values of $\mu_\mathrm{ABS}$. The crossover temperature $T_\mathrm{outer}^-$ is marked for the case $\mu_\mathrm{ABS}=\Delta_0$. Parameters used: $\gamma_1^0=0$ and $\gamma_2^0=0.1\Delta_0$; $t_L=t_R=2/3\Delta_0$.}
    \label{fig:new-entropy}
\end{figure}

\end{widetext}

\end{document}